\documentclass[12pt, draftclsnofoot, onecolumn]{IEEEtran}
\usepackage{cite,graphicx,amsmath,amssymb}
\usepackage{subfigure}
\usepackage{citesort}
\usepackage{fancyhdr}
\usepackage{mdwmath}
\usepackage{mdwtab}
\usepackage{balance}
\usepackage{xcolor}
\usepackage{bm}

\usepackage{algorithm}
\usepackage{algorithmic}

\usepackage{multirow}
\usepackage{flafter}

\usepackage{multicol}
\usepackage{amsthm}
\usepackage{caption}
\usepackage{graphicx}

\newtheorem{remark}{Remark}

\newtheorem{theorem}{Theorem}
\newtheorem{lemma}{Lemma}

\newtheorem{myDef}{Definition}
\newtheorem{myPro1}{Problem}

\begin{document}

%\title{Fairness of User Clustering in MIMO Non-orthogonal Multiple Access Systems}
%\author{Author 1,  Author 2, Author 3, and Author 4}
%%\markboth{IEEE Transactions on \LaTeX\ }
%%{Hayes}

\title{Learning Automata Based Q-learning for Content Placement in Cooperative Caching}
\author{Zhong\ Yang,~\IEEEmembership{Student Member,~IEEE,}
Yuanwei\ Liu,~\IEEEmembership{Member,~IEEE,}\\
Yue\ Chen,~\IEEEmembership{Senior Member,~IEEE,}
Lei\ Jiao,~\IEEEmembership{Senior Member,~IEEE}

\thanks{ Part of this paper was presented in IEEE Global Communication Conference (GLOBECOM), Abu Dhabi, UAE, Dec. 2018~\cite{zhong2018GC}.}
\thanks{ Z. Yang, Y. Liu and Y. Chen are with the School of Electronic Engineering and Computer Science, Queen Mary University of London, London E1 4NS, UK. (email:\{zhong.yang, yuanwei.liu, yue.chen\}@qmul.ac.uk),

L. Jiao is with the Department of Information and Communication Technology, University of Agder, 4879 Grimstad, Norway. (email: lei.jiao@uia.no).}
}
\maketitle

\vspace{-0.8cm}
\begin{abstract}
An optimization problem of content placement in \emph{wireless cooperative caching} is formulated, with the aim of maximizing the sum mean opinion score (MOS) of mobile users. Firstly, as user mobility and content popularity has a significant impact on the system performance, a supervised feed-forward back-propagation connectionist model based neural network (SFBC-NN) is invoked for \emph{user mobility prediction} and \emph{content popularity prediction}. More particularly, practical data collected from GPS-tracker app on smartphones is tackled to test the accuracy of user mobility prediction. Then, based on the predicted mobile users' positions and content popularity, a learning automata based Q-learning (LAQL) algorithm for wireless cooperative caching is proposed, in which learning automata (LA) is invoked for Q-learning to obtain an optimal action selection in a random and stationary environment. It is proven that the LA based action selection scheme is capable of enabling every state to select the optimal action with arbitrarily high probability if Q-learning is able to converge to the optimal Q value eventually. To characterize the performance of the proposed LAQL algorithms, sum MOS of users is applied to define the reward function. Extensive simulation results reveal that: 1) The prediction error of SFBC-NN based algorithm lessen with the increase of iterations and nodes; 2) the proposed LAQL achieves significant performance improvement against traditional Q-learning algorithm; and 3) the cooperative caching scheme is capable of outperforming non-cooperative caching and random caching of $3\% $ and $4\% $.
\end{abstract}

% key words
\vspace{-0.8cm}

\section{Introduction}

It is expected that the first fifth generation (5G)-network based solutions will be commercially launched by 2020~\cite{Montalban2018MCOM}. According to a recent report from Cisco~\cite{Cisco2017}, mobile data traffic has grown 18-fold over the past 5 years and will grow at a compound annual growth rate (CAGR) of 47 percent from 2016 to 2021. More particularly, mobile video traffic accounted for 60 percent of total mobile data traffic in 2016, and over three-fourths (78 percent) of the world's mobile data traffic will be video by 2021. In the vast area of 5{G} wireless systems, an extraordinary variety of technological innovations was included to support a large number of devices over limited spectrum resources~\cite{Yuanwei2018ArXiv,Qin2016TSP,Liu2017PI}. Investigation presented in~\cite{Tombaz2014ICC} shows that backhaul consumes up to 50 percent of the power in a wireless access network. Proactive caching at the base stations (BSs) is a promising approach to dealing with this problem, by introducing cache capabilities at BSs and then pre-fetching contents during off-peak hours before being requested locally by users (see~\cite{Ioannou2016ICST}, and references therein).

Designing proactive caching policy for each BS independently (e.g., each BS caching the most popular contents) may result in insufficient utilization of caching resources (see~\cite{liu2016cm}, and references therein). Cooperative caching increases content diversity in networks by exchanging availability information, thus further exploits the limited storage capacity and achieves more efficient wireless resource utilization (see \cite{Liao2017WCOM,Zhang2017Trans,Wang2017ICL,Liu2017JSAC}, and references therein). The key idea of cooperative caching is to comprehensively utilize the caching capacity of BSs to store specific contents, thus improving QoE of all users in networks. In \cite{Liao2017WCOM}, multicast-aware cooperative caching is developed using maximum distance separable (MDS) codes for minimizing the long-term average backhaul load. In \cite{Zhang2017Trans}, delay-optimal cooperative edge caching is explored, where a greedy content placement algorithm is proposed for reducing the average file transmission delay.

%In \cite{Wang2017ICL}, cooperative caching was considered when BS caching and D2D caching coexist, the content placement problem was formulated to maximize successful transmission probability.

Based on aforementioned advantages of wireless caching, research efforts have been dedicated to caching solutions in wireless communication systems. Aiming to minimize the average downloading latency, the caching problem was formulated as an integer-linear programming problem in~\cite{Jiang2017Deli}, which are solved by subgradient method. In~\cite{Wang2018TWC}, multi-hop cooperative caching is proposed for improving the success probability of content sharing~\cite{Liu2017TWC}. The emerging machine learning paradigm, owing to its broader impact, has profoundly transformed our society, and it has been successfully applied in many applications~\cite{Zhijin2018DL}, including playing Go games~\cite{David2016nature}, playing atari~\cite{Mnih2013arxiv}, human-level control~\cite{Mnih2015nature}. Wireless data traffic contains strong correlations and statistical features in various dimensions, thereby the application of machine learning in wireless cooperative caching networks optimization presents a novel perspective.

\subsection{Motivation and Related Works}

Sparked by the aforementioned potential benefits, we therefore explore the potential performance enhancement brought by reinforcement learning for wireless cooperative caching networks. The authors in \cite{Muller2017TWC} proposed a context-aware proactive caching algorithm, which learns content popularity online by regularly observing context information of connected users. In \cite{li2017jsac}, content caching is considered in a software-defined hyper-cellular networks (SD-HCN), aiming at minimizing the average content provisioning cost.
%In\cite{Ghanavi2018arXiv}, Q-learning based aerial base station (aerial-BS) assisted terrestrial network is considered to provide an effective placement strategy which increases the QoS of wireless networks.
%By integrating deep learning with Q-learning, deep Q-learning or deep Q-network (DQN) utilizes a deep neural network with states as input and the estimated Q values as output, to efficiently learn policies for high-dimensional, large state-space problems~\cite{Mnih2013arxiv}.

In\cite{Ghanavi2018arXiv}, Q-learning based aerial base station (aerial-BS) assisted terrestrial network is considered to provides an effective placement strategy which increases the QoS of wireless networks. By integrating deep learning with Q-learning, deep Q-learning or deep Q-network (DQN) utilize a deep neural network with states as input and the estimated Q values as output, to efficiently learn policies for high-dimensional, large state-space problems~\cite{Mnih2013arxiv}. In \cite{Chen2017TWC}, an approach to perform proactive content caching was proposed based on the powerful frameworks of echo state networks (ESNs) and sublinear algorithms, which predicts the content popularity and users' periodic mobility pattern. In \cite{Chen2017arXiv}, ESNs are used to effectively predict each user's content request distribution and its mobility pattern when limited human-centric information such as users' visited locations, requested contents, gender, job, and device type, etc.

Q-learning enables learning in an unknown environment as well as overcoming the prohibitive computational requirements, which has been widely used in wireless networks~\cite{Somuyiwa2017}. DQNs are used in \cite{Wang2018arXiv} to solve a dynamic multichannel access problem. During peak hours, backhaul links become congested, which makes the QoE of end users low. One approach to mitigate this limitation is to shift the excess load from peak periods to off-peak periods. Caching realizes this shift by fetching popular contents, e.g., reusable video streams, during off peak periods, storing these contents in BSs equipped with memory units, and reusing them during peak traffic hours~\cite{Paschos2016}.

%During peak hours, backhaul links become congested, which makes the QoE of end users low. One approach to mitigate this limitation is to shift the excess load from peak periods to off-peak periods. Caching realizes this shift by fetching popular contents, e.g., reusable video streams, during off peak periods, storing these contents in BSs equipped with memory units, and reusing them during peak traffic hours~\cite{Paschos2016}.

%Q-learning methods use tabular representation (i.e., Q table) to learn the Q value of taking an action from each possible state in order to maximize the long-term reward. For some scenarios with large state spaces and action space, the main handicap of conventional Q-learning is to deal with large number of state and action space. Learning automata \cite{Zhang2016} is an adaptive decision-making method used in the unknown environment, which combines fast and accurate convergence with low computational complexity. In \cite{Kumaravelan2010ICCICR}, the authors addressed the above problem by hierarchical organization of automaton to learn optimal strategy.

\subsection{Contributions and Organization}

While the aforementioned research contributions have laid a solid foundation on wireless cooperative caching, the investigations on the applications of machine learning in wireless cooperative caching are still in their fancy~\cite{Liu2016JSAC1}. In this article, we study the wireless cooperative caching system, where the caching capacity and the backhaul capacity are restricted. It is worth pointing out that the characteristics of cooperative caching make it challenging to apply the reinforcement learning, because the number of caching states increase exponentially with the number of contents and BSs. With the development of reinforcement learning and high computing speed of new computer, we design an autonomous agent, that perceives the environment, and the task of the agent is to learn from this non-direct and delayed payoff so as to maximize the cumulative effect of subsequent actions. The agent improves its performance and chooses behavior through learning.

Driven by solving all the aforementioned issues, we present a systematic approach for content cooperative caching. More specifically, we propose a learning automata based Q-learning (LAQL) algorithm for content placement in wireless cooperative caching networks (i.e., we consider ``what'' and ``where" to cache). Instead of complex calculation of the optimal content placement, our designed algorithm enables a natural learning paradigm by state-action interaction with the environment it operates in, which can be used to mimic experienced operators.  Our main contributions are summarized as follows.

\begin{enumerate}
  \item We propose a content cooperative caching framework to improve QoE of users. We establish the correlation between user mobility, content popularity, and content placement in cooperative caching, by modeling the content placement with user mobility and content popularity. Then, we formulate the sum MOS maximization problem subject to the BSs' caching capacity, which is a combinatorial optimization problem. We mathematically prove that the formulated problem is nondeterministic polynomial-time (NP) hard.

  \item We invoke a supervised feed-forward back propagation connectionist models based neural network (SFBC-NN) for user mobility prediction and content popularity prediction. We explore the number of nodes and iterations of the SFBC-NN to improve the prediction accuracy. We utilize practical trajectory data collected from GPS-tracker app on smartphones to validate the accuracy of user mobility prediction\footnote{The dataset has been shared by authors in Github. It is shown on the websit: https://github.com/swfo/User-Mobility-Dataset.git. Our approach can accommodate other datasets without loss of generality.}.

  \item We conceive an LAQL algorithm based solution for content placement in cooperative caching. In contrast to a conventional Q-learning algorithm, we design an action selection scheme invoking learning automata (LA). Additionally, we prove that the LA based action selection scheme is capable of enabling every state to select the optimal action with arbitrary high probability if Q-learning is capable of converging to the optimal Q value eventually.

  \item We demonstrate that the performance of the proposed LAQL based content cooperative caching with content popularity and user mobility prediction outperforms conventional Q-learning. Meanwhile content cooperative caching is capable of outperforming non-cooperative caching and random caching algorithm in terms of QoE of users.

\end{enumerate}

The rest of the paper is organized as follows. In Section~\ref{section:SystemModel}, the system model for content caching in different BSs is presented. In Section~\ref{section:ANNbasedUMP}, the SFBC-NN based user mobility and content popularity prediction are investigated. LAQL for content cooperative caching is formulated in Section~\ref{section:SystemModelandProblemFormulation}. Simulation results are presented in Section~\ref{section:Simulationresults}, before we conclude this work in Section~\ref{section:Conclusion}. Table~\ref{tablenotation} provides a summary of the notations used in this paper.

\begin{table*}[h]
	\caption{LIST OF NOTATIONS}
	\centering
	\begin{tabular}{|c|c|c|c|}\hline
		Notation&Description&Notation&Description\\\hline
        $M$ & The number of BSs& $R_{m\min }$ & The minimum fronthaul capacity of BS $m$  \\\hline
        $N_u$ & The number of users& ${\rm{MO}}{{\rm{S}}_i}(t)$ & The mean opinion score of user $i$ at time $t$  \\\hline
		$F$ & The umber of contents& ${\rm RTT}$ & The round trip time  \\\hline
        $l\left( {} \right)$ & The size of content & ${\rm FS}$ & The web page size  \\\hline
        ${z_i}$ & Position of user $i$ & ${\rm MSS}$ & The maximum segment size  \\\hline
        ${z_m}$ & Position of BS $m$ & $L$ & The number of slow start cycles with idle periods  \\\hline
        ${s_m}$ & Caching capacity of BS $m$ & ${x_{l - 1}^{input}}$ & The input of layer $l-1$  \\\hline
        ${X_m}$ & Caching vector of BS $m$ & ${{W^l}}$ & The weights of layer $l$  \\\hline
        ${r_{mi}}$ & Distance between user $i$ and BS $m$ & ${{b^l}}$ & The bias of layer $l$  \\\hline
        $\rho $ & Transmit power of BSs & ${{y^{output}}}$ & The output of the network  \\\hline
        ${{h_{mi}}}$ & Small scale fading of BS $m$ and user $i$ & ${{y^{real}}}$ & The real data  \\\hline
        ${{\sigma ^2}}$ & The power of the additive noise & ${Q^\pi }\left( {s,a} \right)$ & The Q table  \\\hline
        ${{I_i}}$ & The sum of interfering signal power & ${Q^*}\left( {s,a} \right)$ & The optimal state-action value function  \\\hline
        $B$ & The bandwidth of each downlink user & ${u_i}\left( t \right)$ & The times action $i$ has been rewarded when selected  \\\hline
        ${{p_{t,f}}}$ & The popularity of content $f$ at time $t$ & ${v_i}\left( t \right)$ & The number of times that action $i$ has been selected  \\\hline
        $R_{m\max }$ & The maximum fronthaul capacity of BS $m$ & ${P_a}\left( t \right)$ & The probability distribution of LA  \\\hline

     %   $R_{m\min }$ & The minimum fronthaul capacity of BS $m$
%        ${\rm{MO}}{{\rm{S}}_i}(t)$ & The mean opinion score of user $i$ at time $t$
%        ${\rm RTT}$ & The round trip time
%        ${\rm FS}$ & The web page size
%        ${\rm MSS}$ & The maximum segment size
%        $L$ & The number of slow start cycles with idle periods
%        ${x_{l - 1}^{input}}$ & The input of layer $l-1$
%        ${{W^l}}$ & The weights of layer $l$
%        ${{b^l}}$ & The bias of layer $l$
%        ${{y^{output}}}$ & The output of the network
%        ${{y^{real}}}$ & The real data
%        ${Q^\pi }\left( {s,a} \right)$ & The Q table
%        ${Q^*}\left( {s,a} \right)$ & The optimal state-action value function
%        ${u_i}\left( t \right)$ & The number of times that action $i$ has been rewarded when selected
%        ${v_i}\left( t \right)$ & The number of times that action $i$ has been selected
%        ${P_a}\left( t \right)$ & The probability distribution of LA
%
      \end{tabular}
	\label{tablenotation}
\end{table*}

\section{System Model}\label{section:SystemModel}

\subsection{System Description}

We consider a square cooperative area composed of $M$ base stations (BSs) with one transmit antennas and $N_u$ single antenna users (as is shown in Fig.~\ref{scenariopic}). The central processor collects relevant information from the system, and acts as a master to control the BS (slaves), and implements a centrally controlled adjustment of caching policies~\cite{Sadeghi2018JSAC}. Denote $\mathcal{M} = \left\{ {1, \cdots ,M} \right\}$ and $\mathcal{N} = \left\{ {1, \cdots ,N_u} \right\}$ be the BSs set and the user set, respectively. We assume that the content library in the central network consists of $F$ contents denoted by $\mathcal{F} = \left\{ {1, \cdots ,F} \right\}$, and the size of $f$-th content is $l\left( f \right)$. The position of user $i$ is denoted as $z_i$. The position of BS $m$ is denoted by $z_m$, and the caching capacity of BS $m$ is $s_m$. The content cache vector of BS $m$ is ${X_m} = \left[ {{x_{1,m}},{x_{2,m}} \cdots ,{x_{{s_m},m}}} \right]$, where ${x_{i,m}} \in \left[ {1,F} \right]$. Therefore, the cooperative caching matrix is denoted as:

\begin{figure} [t!]
 \centering
 \includegraphics[width=3.5in]{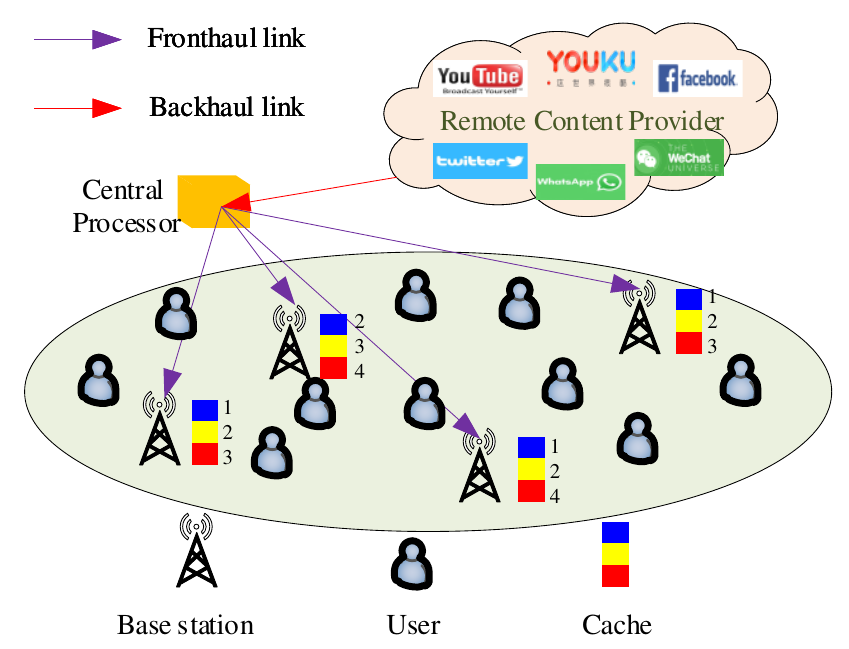}
 \centering
 \caption{Schematic of the wireless cooperative caching scenario. The users are served by BSs cooperatively, and the contents cached in different BSs (slaves) are placed by cental processor (master).}\label{scenariopic}
\end{figure}

\begin{equation}\label{cachingmetrix}
X = \left[ {\begin{array}{*{20}{c}}
{{X_1}}\\
{{X_2}}\\
 \vdots \\
{{X_M}}
\end{array}} \right] = \left[ {\begin{array}{*{20}{c}}
{{x_{1,1}},{x_{2,1}} \cdots ,{x_{{s_1},1}}}\\
{{x_{1,2}},{x_{2,2}} \cdots ,{x_{{s_2},2}}}\\
 \vdots \\
{{x_{1,M}},{x_{2,M}} \cdots ,{x_{{s_M},M}}}
\end{array}} \right].
\end{equation}

We set the position of users randomly, and the position of user $i$ is denoted by ${z_i} = \left( {{x_i},{y_i}} \right)$, and the position of BS $m$ is denote by ${z_m} = \left( {{x_m},{y_m}} \right)$. As such, the distance from user $i$ to BS $m$ is given by $\left\| {{r_{mi}}} \right\| = \left\| {{z_i} - {z_m}} \right\| = \sqrt {{{\left( {{x_i} - {x_m}} \right)}^2} + {{\left( {{y_i} - {y_m}} \right)}^2}} $. We consider a wireless communication system with FDMA.

Consider content $f$ in the cooperative region, which is cached in multiple BSs, denoted by $\mathcal{M}_f$. These BSs jointly transmit content $f$ to user $i$ who requests for it. The received signal-interference-noise ratio (SINR) of received signal of user $i$ requesting for content $f$ is given by

%The received signal of user $i$ is given by
%
%\begin{equation}\label{signal}
%{y_i^{receive} = \underbrace {\sum\limits_{m \in \left\{ {{\mathcal{M}_f}} \right\}} {\sqrt \rho  {h_{mi}}{{\left| {{r_{mi}}} \right|}^{ - \alpha /2}}s} }_{{\rm{Signal}}} + \underbrace {{\sum _{n \in \left\{ \mathcal{M} \right\}\backslash \left\{ {{\mathcal{M}_f}} \right\}}}\sqrt \rho  {h_{ni}}{{\left| {{r_{ni}}} \right|}^{ - \alpha /2}}{s_n}}_{{\rm{Interference}}} + n,}
%\end{equation}
%where ${\rho}$ denotes the equally transmit power of all BSs, and ${\left| {{r_{mi}}} \right|^{ - \alpha }}$ represents a standard distance-dependent power law pathloss attenuation between BS $m$ and user $i$, where $\alpha  \ge 2$ is the pathloss exponent. ${h_{mi}}$ denotes the small-scale Rayleigh fading between BS $m$ and user $i$. And $n$ denotes a standard additive white Gaussian noise.

\begin{equation}\label{SINR}
{\rm{SIN}}{{\rm{R}}_{mi}} = \frac{{{{\left( {\sum\limits_{m \in \left\{ {{\mathcal{M}_f}} \right\}} {\sqrt \rho  \left| {{h_{mi}}} \right|r_{mi}^{ - \alpha /2}} } \right)}^2}}}{{{I_i} + {\sigma ^2}}},
\end{equation}
where ${\rho}$ denotes the equally transmit power of all BSs, and ${\left| {{r_{mi}}} \right|^{ - \alpha }}$ represents a standard distance-dependent power law pathloss attenuation between BS $m$ and user $i$, where $\alpha  \ge 2$ is the pathloss exponent. ${h_{mi}}$ denotes the small-scale Rayleigh fading between BS $m$ and user $i$, ${{\sigma ^2}}$ denotes the power of the additive noise and $I_i$ is the sum of interfering signal power from BSs that do not cache content $f$ given by

\begin{equation}\label{interference}
{I_i} = {\sum _{n \in \left\{ \mathcal{M} \right\}\backslash \left\{ {{\mathcal{M}_f}} \right\}}}\rho {\left| {{h_{ni}}} \right|^2}r_{ni}^{ - \alpha },
\end{equation}

The bandwidth of each downlink user is denoted as $B$, Therefore, based on Shannon's capacity formula, the overall achievable sum rate of user $i$ at time $t$ can be expressed as

\begin{equation}\label{overall achievable sum rate}
{R_{t,i}} = B\sum\limits_{f = 1}^F {\left( {{p_{t,f}}\sum\limits_{m = 1}^M {{x_{t,mf}}{{\log }_2}\left( {1 + SIN{R_{t,im}}} \right)} } \right)} .
\end{equation}
where $p_{t,if}$ denotes content $f$'s popularity at time $t$.

If the content is not cached in the BS, then the content needs to be fetched from the centre network using fronthaul link. Due to the fronthaul link capacity constrains, the total transmitting rate of each BS should not exceed its fronthaul capacity $R_{m\max }$. Assuming there are $H_m$ users associated with BS $m$, thus

\begin{equation}\label{fronthaul capacity constrain}
B\sum\limits_{i = 1}^{{H_m}} {\left( {{p_{t,mi}}{x_{t,mf}}{{\log }_2}\left( {1 + SIN{R_{t,im}}} \right)} \right)}  \le {R_{m\max }}.
\end{equation}

Another constrain is that the transmit rate for each user should not be less than the minimum transmit rate ${R_{i\min }}$, thus

\begin{equation}\label{transmit rate constrain}
\sum\limits_{m = 1}^M {\left( {{p_{t,mi}}{x_{t,mf}}{{\log }_2}\left( {1 + SIN{R_{t,im}}} \right)} \right)}  \ge {R_{i\min }}.
\end{equation}

\subsection{Quality-of-Experience Model}

As in \cite{itu2016qoe}, the definition of quality-of-experience (QoE) is ``The overall acceptability of an application or service, as perceived subjectively by the end user." the QoE is a kind of users' satisfaction description during the process of interactions between users and services~\cite{wang2017qoe}. Based on~\cite{Rugelj2014TC}, we define our MOS of user $i$ in time $t$ as

\begin{equation}\label{MOS1}
{\rm{MO}}{{\rm{S}}_i}(t) =  - {C_1}\ln \left( {d\left( {{R_{t,i}}} \right)} \right) + {C_2},
\end{equation}
where $C_1$ and $C_2$ are constants determined by analyzing the experimental results of the web browsing applicants, which are set to be 1.120 and 4.6746, respectively. ${\rm{MO}}{{\rm{S}}_i}(t)$ represents the user perceived quality expressed in real numbers ranging from 1 to 5 (i.e., the score 1 represents extremely low quality, whereas score 5 represents excellent quality. $R_{t,i}$ is achievable sum rate of user $i$ at time $t$, and ${d\left( {{R_{t,i}}} \right)}$ is the delay time of a user receiving a content. According to ~\cite{ameigeiras2010qoe},

\begin{equation}\label{distRsum}
{d\left( {{R_{t,i}}} \right) = 3{\rm{RTT}} + \frac{{{\rm{FS}}}}{{{R_{t,i}}}} + L\left( {\frac{{{\rm{MSS}}}}{{{R_{t,i}}}} + {\rm{RTT}}} \right) - \frac{{2{\rm{MSS}}\left( {{{\rm{2}}^{\rm{L}}}{\rm{ - 1}}} \right)}}{{{R_{t,i}}}}},
\end{equation}
where ${R_{t,i}}$ represents the data rate, ${\rm RTT}$ is the round trip time, ${\rm FS}$ is the web page size, and ${\rm MSS}$ is the maximum segment size. $L$ is the number of slow start cycles with idle periods, $L = \min \left[ {{L_1},{L_2}} \right]$, which is defined as~\cite{ameigeiras2010qoe}: ${L_1} = {\log _2}\left( {\frac{{{R_{t,i}}{\rm{RTT}}}}{{{\rm{MSS}}}} + 1} \right) - 1$, ${L_2} = \log _2\left( {\frac{{{\rm FS}}}{{2{\rm MSS}}} + 1} \right) - 1$, where $L_1$ denotes the number of cycles that the congestion window takes to reach the bandwidth-delay product and $L_2$ is the number of slow start cycles before the webpage size is completely transferred.

Substituting (\ref{distRsum}) into (\ref{MOS1}), we obtain MOS of user $i$ as follows:

\begin{equation}\label{MOSall}
{\rm{MO}}{{\rm{S}}_i}(t) =  - {C_1}\ln \left[ {3{\rm{RTT}} + \frac{{{\rm{FS}}}}{{{R_{t,i}}}}\left. { + L\left( {\frac{{{\rm{MSS}}}}{{{R_{t,i}}}} + {\rm{RTT}}} \right) - \frac{{2{\rm{MSS}}\left( {{{\rm{2}}^L}{\rm{ - 1}}} \right)}}{{{R_{t,i}}}}} \right] + {C_2}.} \right.
\end{equation}

%\begin{figure} [t!]
% \centering
% \includegraphics[width=3.5in]{eps/Flowchartforcaching.eps}
% \centering
% \caption{An overview of proposed framework for cooperative caching.}\label{problemtransformation}
%\end{figure}

\section{SFBC-NN Based User Mobility and Content Popularity Prediction}\label{section:ANNbasedUMP}

\begin{figure} [t!]
 \centering
 \includegraphics[width=3.5in]{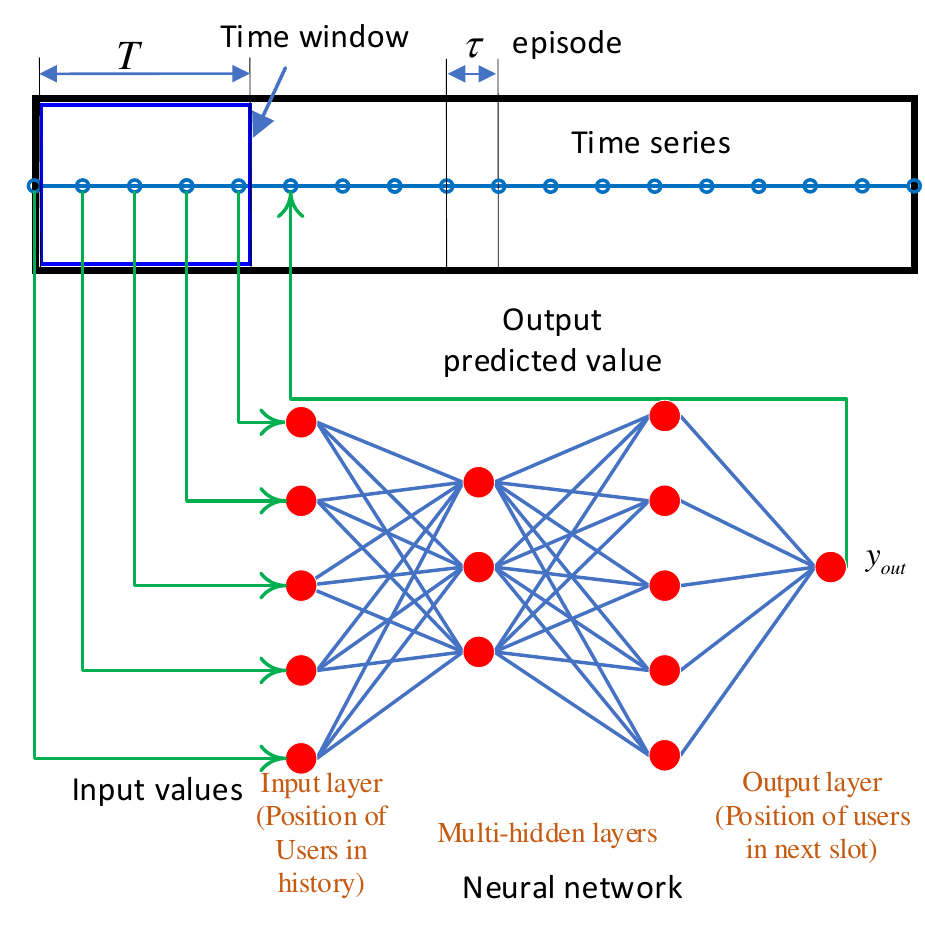}
 \centering
 \caption{Flow chart of the SFBC-NN for time series prediction.}\label{flowchartforneutalnetwork}
\end{figure}

%For time series prediction problem, let the observation window of history information be over the last $K$ time slots. In time $t$, let ${x_t}$ be the value of time series. The object of time series prediction is to find a prediction function ${x_{t + 1}} = f\left( {{x_{t - K + 1}},{x_{t - K + 2}}, \cdots {x_t}} \right)$ that achieves ~\cite{Qiu2018WCLetter}:
%
%\begin{equation}\label{predictionproblem}
%\mathop {\min }\limits_f \mathop {\lim }\limits_T \frac{1}{T}\sum\limits_{t = 1}^T {L\left( {{{\widehat x}_{t + 1}},{x_{t + 1}}} \right)},
%\end{equation}
%where the loss function $L\left(  \bullet  \right)$ denotes the loss between the real value and the predicted value. According to~\cite{Qiu2018WCLetter}, a general solution for (\ref{predictionproblem}) is intractable. Therefore we propose a neural network for the above problem.

%\subsection{User Mobility Prediction}

In reality, users are mobile, and may have periodic mobility patterns. We assume that the time interval of the change of users' mobility is $T$. Universal approximation theorems in~\cite{HORNIK1991251} show that neural networks have a striking fact that they compute any function at all. Hencefore in this section, we propose an SFBC-NN based user position and content popularity prediction algorithm. The structure of the SFBC-NN for user mobility prediction contains two hidden layers and a fully connected output layer. The inputs are users' positions in one hour containing 12 positions.

\begin{myDef}\label{Def.SFBC-NN}
The definition of three parameters adopted in SFBC-NN are given below.
\begin{itemize}
  \item Input layer (IL): Passive layer, receiving single value of input, and duplicating the value to multiple outputs. ${x_{t,i}}$ represents the position of the $i$-th user at time $t$, and the network input is the history combination of $M$ positions, i.e., $x_{}^{input} = \left[ {{x_{t - M,i}},{x_{t - M + 1,i}}, \cdots ,{x_{t - 1,i}}} \right]$. Here we set $M=12$ which means that we use history one hour positions to predict positions in the next hour.
%  \item Hidden Layer (HL): Active layer, modifying the data of input, and producing an output through an activation function (mostly sigmoid or tanh). There are two hidden layers in our model,(i.e., HL1 and HL2).
%  The output of hidden layer is connected to the inputs of other neurons and therefore is not visible as a network output.
  \item Output Layer (OL): Active layer, modifying the data of input, and exporting given outputs for the network. $y_{}^{output} = \left[ {{y_{t + 1,i}},{y_{t + 2,i}}, \cdots ,{y_{t + N,i}}} \right]$ represents the positions of the $i$-th user for the next step, where $N$ represents length of the track.
  \item Root mean square error (RMSE): The performance of the SFBC-NN based user mobility prediction developed in our work is evaluated by RMSE, which adjusts the weight of the network after each iteration, thus it affects the rate of convergence. The SFBC-NN produces a function relation between input ${I_{t,if}}$ and output ${O_{t,if}}$:

  \begin{equation}\label{transforequation}
y_l^{output} = f\left( {{W^l}x_{l - 1}^{input} + {b^l}} \right),
  \end{equation}
  where $W^l$ denotes the weight matrix of layer $l$, and $b^l$ denotes the bias of layer $l$. $f\left(  \bullet  \right)$ is the activation function (i.e., Sigmoid Function.), which is applied to node inputs to produce node output. The forward propagation algorithm will pass the information through the network to make a prediction in the output layer. After forward propagation, as in \textbf{Definition~\ref{Def.SFBC-NN}},  the network uses a RMSE function to measure the loss between trained output and output:

  \begin{equation}\label{lossfunction}
{\rm RMSE}\left( {{y^{output}},{y^{real}}} \right) = \frac{1}{{{N_u}}}\sum\limits_{i = 1}^{{N_u}} {\sqrt {\frac{1}{{{N_o}}}\sum\limits_{j = 1}^{{N_o}} {{{\left[ {y_i^{output}\left( j \right) - y_i^{real}\left( j \right)} \right]}^2}} } } ,
  \end{equation}
  where ${{y^{output}}}$ and ${{y^{real}}}$ denotes the output of the network and real data, respectively. The details of SFBC-NN based user mobility prediction are shown in \textbf{Algorithm}~\textbf{\ref{cpprediction}}.
\end{itemize}
\end{myDef}

A typical SFBC-NN architecture contains input (denoted by $X_t$), the hidden layers ($H$), and output layers (denoted by $Y$). In our SFBC-NN, we desire a double hidden layers SFBC-NN to predict user mobility.

\begin{algorithm}[h]
\caption{The SFBC-NN for user mobility prediction.}
\label{cpprediction}
\begin{algorithmic}[1]

    \STATE \textbf{Stage One: Training }
\REQUIRE Time interval $H$, training data: positions of users $\left[ {{x_{t - H + 1,i}},{x_{t - H + 2,i}}, \cdots ,{x_{t,i}}} \right]$. Parameters of SFBC-NN: number of hidden layers 2, learning rate 0.1, goal of RMSE, the number of epochs.

    \STATE Randomly initialize number of hidden layers, the weight matrix $W^l$ and the bias $b^l$.
    \FOR{each episode}
    \STATE Choose input $I_{t,if}$ from training set;\
    \STATE Forward propagation algorithm: calculating the output of the layers ${O^l}(l = 1:{N_{ne}})$;\
    \STATE Calculate loss function $los{s^l}(l = 1:{N_{ne}})$;\
    \STATE Backward propagation algorithm: adjust the weights ${W^l}(l = 1:{N_{ne}})$ and bias ${b^l}(l = 1:{N_{ne}})$ according to the loss function;\
    \ENDFOR
    \ENSURE Weights ${W^l}(l = 1:{N_{ne}})$ and bias ${b^l}(l = 1:{N_{ne}})$ of the network.\

    \STATE \textbf{Stage Two: Testing}
\REQUIRE Interval $H$, Testing data: Positions of users, Parameters of SFBC-NN: weights ${W^l}(l = 1:{N_{ne}})$, bias ${b^l}(l = 1:{N_{ne}})$ of the network, the number of epochs.

    \FOR{each episode}
    \STATE Choose input $I_{t,if}$ from testing set;\
    \STATE Forward propagation: calculating the output of the layers ${O^l}(l = 1:{N_{ne}})$;\
    \STATE Calculate loss function $los{s^l}(l = 1:{N_{ne}})$;\
    \ENDFOR
    \ENSURE The accuracy between predicted user's positions and real data.
\end{algorithmic}
\end{algorithm}

%One challenge in SFBC-NN design is that the selection of the optimal number of units, which should be large enough to fit the purpose. Nevertheless, the number of units cannot be too large, which makes the SFBC-NN fails to overfitting~\cite{Wanjawa2015arXiv}.

%There are several issues to consider in the design of SFBC-NN, which are:
%
%\begin{itemize}
%  \item The type of network: In our SFBC-NN network, we first use forward propagation to calculate the output of the network, then we use backward propagation to adjust the weight matrix of the layers;
%  \item The type of training: We use supervised training to train the network; When we train the network, the network gets weights and biases that find the relevant patterns to make better predictions;
%  \item The number of input and output units: Taking a larger number of inputs yields accurate prediction but incurs a higher network complexity. Waiting too long to declare the predicted mobility leads to delay. In this work we use the formal 12 points to predict the 13th point, so the number of inputs is 12;
%  \item The number of hidden layers: We design a SFBC-NN network which contains 2 hidden layers.
%\end{itemize}

\subsection{Convergence of SFBC-NN}

We obtain local optimal solution with gradient method training. Better training algorithms for SFBC-NN is needed to obtain global optimal solution, because it is more representationally efficient than shallow ones~\cite{bengio2007greedy}.

\begin{remark}\label{remark:Convrgence of SFBC-NN}
For SFBC-NN based user mobility prediction, layer-wise pre-training (LWPT) is generally adopted, which means that in the beginning of the whole network, only a small number of separate neurons need to be trained, then the whole network is trained by gradient. In each step, we fix the pre-trained $k-1$ layers and add the $k$-th layer (that is, the input of $k$-th layer is the output of the former $k-1$ layers that we have trained). Each level of training is supervised (i.e., taking the prediction error of each step as the loss function).
\end{remark}

\subsection{Stability of SFBC-NN}

The stability of SFBC-NN with respect to small perturbations of the inputs is meaningful, because small perturbations like users' unfrequent visit positions and abnormal contents surfing are not main concern in our scenario. We focus on scenarios that people periodical visit and contents that they request stably in the areas.

\begin{remark}\label{remark:Stability of SFBC-NN}
Results suggest that the SFBC-NN for user mobility prediction that are learned by backpropagation have nonintuitive characteristics and intrinsic blind spots, and its structure is connected to the data distribution in a non-obvious way~\cite{szegedy2013intriguing}. Thus in our scenario, data preprocessing (i.e., perturbations elimination) is of great importance to increase the stability of SFBC-NN based user mobility prediction.
\end{remark}

\begin{remark}\label{remark:DRL}
An apparent advantage of utilizing the SFBC-NN for content popularity prediction is that it stores the non-linear characters in the multi-parameters of SFBC-NN, which reflects the nature of the relationship between the inputs and the outputs.
\end{remark}

In (\ref{overall achievable sum rate}), content popularity $p_{t,mi}$ plays an important role in overall achievable sum rate, therefor content popularity is another element that we predict for content cooperative caching. Note that the dimension of content popularity is larger than user mobility. For this reason, we set more hidden layers than user mobility prediction network. The structure of the neural network based content popularity prediction contains three hidden layers and a fully connected output layer. The input data is partitioned into blocks of size ${k_{in}} = 5 \times 1$, where each block represents for content popularity in 5 time slots ${X_{in}} = \left[ {{p_{ik}}, \cdots ,{p_{i(k + 4)}}} \right]{\kern 1pt} {\kern 1pt} {\kern 1pt} {\kern 1pt} {\kern 1pt} {\kern 1pt} {\kern 1pt} {\kern 1pt} {\kern 1pt} {\kern 1pt} {\kern 1pt} {\kern 1pt} (k = 1,2, \cdots )$. The output is a block of size ${k_{out}} = 1 \times 1$, where each block represents the content popularity in the next time slot ${Y_{out}} = {p_{i(k + 5)}}{\kern 1pt} {\kern 1pt} {\kern 1pt} {\kern 1pt} {\kern 1pt} {\kern 1pt} {\kern 1pt} {\kern 1pt} {\kern 1pt} (k = 1,2, \cdots )$.

%The predictor can be designed by a mapping function $\Upsilon \left(  \bullet  \right)$ given by
%
%\begin{equation}\label{mapping}
%   {Y_{out}} = \Upsilon \left( {{X_{in}}\left| {(W,b} \right.)} \right).
%\end{equation}

\section{Learning Automata based Q-Learning for Content Cooperative Caching}\label{section:SystemModelandProblemFormulation}

%As is shown in (\ref{SINR}), the mobility of users changes the relative position between BSs and users, thus makes significant influence on the SINR of users. Also, in (\ref{overall achievable sum rate}), content popularity has great power of producing an effect the SINR of users. The nature of influence of user mobility prediction and content popularity prediction on content cooperative caching is still not clear. Therefore in our scenario, research on user mobility prediction and content popularity prediction is investigated. Based on the analysis before, we formulate the mobility prediction and content popularity prediction as time series prediction problem. Meanwhile, we utilize neural network to solve this problem, because they are able to catch hidden and strongly non-linear dependencies between the training set. After which, we propose a automata Q-learning based content cooperative caching algorithm, which consists of state space, action space and reward. Regarding state space and action space, content caching matrix and its update are set as states and actions, for reward, MOS of users is applied to define reward of Q-learning.

\subsection{Problem Formulation}

We obtain users mobility and content popularity from Section \ref{section:ANNbasedUMP}. Hereafter, the objective is to maximize the overall ${\rm{MO}}{{\rm{S}}_{tot}}(t) = \sum\limits_{u = 1}^N {{\rm{MO}}{{\rm{S}}_u}(t)} $ with constrains from content cache vector $\textbf{X}$, content characteristics, and caching capacity of BSs. Let ${G_{t,mf}} = {x_{t,mf}}{\log _2}\left( {1 + \rm{SIN{R}_{t,im}}} \right)$ for similarity. The content caching problem is to determine the cached data files' distribution for each BS aiming to maximize the MOS of users, subject to the cache capacity constraint. The optimization problem is formally expressed as

\begin{subequations}\label{optimizationproblem}

\begin{align}
& \left( \textbf{{P1}} \right) \mathop {{\rm{max}}}\limits_{\textbf{X}} {\rm{MO}}{{\rm{S}}_{tot}}\left( t \right) = \sum\limits_{i = 1}^N {{\rm{MO}}{{\rm{S}}_i}\left( t \right)},\label{optimization_problem}\\
\mbox{s.t.} \quad
& C_1:\;{x_{t,mf}} \in \left\{ {0,1} \right\},\forall f \in F,m \in M,\label{c1}\\
& C_2:\;\sum\limits_{f = 1}^F {{x_{t,mf}}{l_f}}  \le {S_m},\forall t \in T,m \in M,\label{c2}\\
& C_3:\;{p_{t,if}} \in \left( {0,1} \right),\forall f \in F,i \in N_u,\label{c3}\\
& C_4:\;\sum\limits_{m = 1}^M {{s_m}}  \le \sum\limits_{f = 1}^F {{l_f}} ,\label{c4}\\
& C_5:\;\sum\limits_{i = 1}^{{H_m}} {B\left( {{p_{t,mi}}{G_{t,mf}}} \right)}  \le {R_{m\max }},\forall t \in T, \label{c5}\\
& C_6:\;\sum\limits_{m = 1}^M {\left( {{p_{t,mi}}\sum\limits_{m = 1}^M {{G_{t,mf}}} } \right)}  \ge {R_{i\min }},\forall t \in T,\label{c6}
\end{align}
where ${x_{t,mf}}$ denotes the caching decision of BS $m$ and content $f$ at time $t$, ${p_{t,if}}$ denotes the popularity of user $i$ for content $f$ at time $t$, and $N_u$ denotes the number of users. (\ref{c2}) states that the total size of contents cached in BS $m$ should not exceed the content caching capacity of the BS. For simplicity, we assume that each content has the same size 1, i.e., ${l_f} = 1$. This assumption is easily found by dividing contents into blocks with the same size~\cite{Shanmugam2013tit}.  (\ref{c4}) ensures that the total caching capacity of all BSs is smaller than the total size of contents, otherwise the BSs would cache all the contents and the research is not meaningful. (\ref{c5}) and (\ref{c6}) are BS fronthaul link capacity constrain and user minimum transmit rate constrain, respectively.

%\begin{equation}\label{optimization_problem}
%\mathop {{\rm{maxmize}}}\limits_{\textbf{X}} {\rm{MO}}{{\rm{S}}_{tot}}\left( t \right) = \sum\limits_{i = 1}^N {{\rm{MO}}{{\rm{S}}_u}\left( t \right)} ,
%\end{equation}
%
%\vspace{-0.3cm}
%\begin{equation}\label{c1}
%\begin{array}{*{20}{c}}
%{s.t.}&{{x_{t,mf}} \in \left\{ {0,1} \right\},}&{\forall f \in F,m \in M,}
%\end{array}
%\end{equation}
%
%\vspace{-0.3cm}
%\begin{equation}\label{c2}
%\begin{array}{*{20}{c}}
%{\sum\limits_{f = 1}^F {{x_{t,mf}}{l_f}}  \le {S_m},}&{\forall t \in T,m \in M,}
%\end{array}
%\end{equation}
%
%\vspace{-0.3cm}
%\begin{equation}\label{c3}
%\begin{array}{*{20}{c}}
%{{p_{t,if}} \in \left[ {0,1} \right],}&{\forall f \in F,i \in N,}
%\end{array}
%\end{equation}
%
%\vspace{-0.3cm}
%\begin{equation}\label{c4}
%\sum\limits_{m = 1}^M {{s_m}}  \le \sum\limits_{f = 1}^F {{l_f}} ,
%\end{equation}
%
%\vspace{-0.3cm}
%\begin{equation}\label{c5}
%\begin{array}{*{20}{c}}
%{\sum\limits_{i = 1}^{{H_m}} {B\left( {{p_{t,mi}}{G_{t,mf}}} \right)}  \le {R_{m\max }},}&{\forall t \in T,}
%\end{array}
%\end{equation}
%\vspace{-0.3cm}
%\begin{equation}\label{c6}
%\begin{array}{*{20}{c}}
%{\sum\limits_{m = 1}^M {\left( {{p_{t,mi}}\sum\limits_{m = 1}^M {{G_{t,mf}}} } \right)}  \ge {R_{i\min }},}&{\forall t \in T,}
%\end{array}
%\end{equation}
\end{subequations}

\begin{lemma}\label{theorem:optimallemma2}
The optimization problem in (\ref{optimizationproblem}) is a NP-hard problem.

\begin{IEEEproof}
See Appendix A~.
\end{IEEEproof}
\end{lemma}

As indicated by~\textbf{Lemma~\ref{theorem:optimallemma2}}, the problem is NP-hard and therefore the problem has no polynomial-time solution discovered so far. Instead of utilizing conventional optimization approaches, we formulate the problem as a finite Markov decision process (MDP), hereafter, we utilize Q-learning algorithm to solve the problem. In order to improve the performance of Q-learning, we adopt learning automata (LA) for the action selection of every state in Q-learning. We apply the predicted results of user mobility and content popularity from the proposed SFBC-NN algorithm in Section~\ref{section:ANNbasedUMP} as inputs for solving problem in (\ref{optimizationproblem}). User position in (\ref{SINR}) and content popularity in (\ref{overall achievable sum rate}) are predicted by neural network, therefor the solution for (\ref{optimizationproblem}) is content caching matrix.

%\subsection{Automata Q-learning Based cooperative Caching}

%Under policy $\pi$, the performance of cooperative caching, also called the state-action value function is
%
%\begin{equation}\label{pi1}
%\begin{array}{*{20}{l}}
%{{Q^\pi }\left( {s,a} \right) = E\left[ {{r_{t + 1}} + \gamma {r_{t + 2}} + {\gamma ^2}{r_{t + 3}} +  \cdots |s,a} \right]}\\
%{ = E\left[ {{r_{t + 1}} + \gamma \left[ {{r_{t + 1 + 1}} + {\gamma ^{1 + 1}}{r_{t + 1 + 2}} + {\gamma ^{1 + 2}}{r_{t + 1 + 3}} +  \cdots } \right]|s,a} \right]}\\
%{ = {E_{s'}}\left[ {{r_{t + 1}} + \gamma {Q^\pi }\left( {s',a'} \right)|s,a} \right]},
%\end{array}
%\end{equation}
%where ${{R_{t + 1}}}$ denotes the reward in the next time slot. ${s'}$ and ${a'}$ denote the state and action in the next time slot.
\subsection{Q-learning Based Solution}
Q-learning is an off-policy and model-free reinforcement learning algorithm compared to policy iteration and value iteration. Moreover, the agent of Q-learning does not need to traverse all states and actions on contrary to value iteration. Q-learning stores state and action information into a Q table ${Q^\pi }\left( {s,a} \right)$, where $\pi$ is the policy, $s$ represents the state, and $a$ represents the action. The objective of the Q-learning is to find optimal policy ${\pi^*}$, and estimate the optimal state-action value function ${Q^*}\left( {s,a} \right): = {Q^\pi }\left( {s,a} \right),\forall s,a$, which can be shown that~\cite{Sutton2016rl}

 \begin{equation}\label{pi2}
{\pi ^*}\left( s \right) = \mathop {\arg \max }\limits_{{\kern 1pt} {\kern 1pt} {\kern 1pt} {\kern 1pt} {\kern 1pt} {\kern 1pt} {\kern 1pt} {\kern 1pt} {\kern 1pt} {\kern 1pt} {\kern 1pt} {\kern 1pt} {\kern 1pt} {\kern 1pt} {\kern 1pt} {\kern 1pt} {\kern 1pt} {\kern 1pt} {\kern 1pt} {\kern 1pt} a} {Q^*}\left( {s,a} \right),\forall s \in S,
\end{equation}
where ${\pi^*}\left( s \right)$ denotes the optimal policy of state $s$, and $S$ denotes the set of all states.

The expression for optimal strategy ${V_{{\pi ^ * }}}\left( s \right)$ is rewritten as

\begin{equation}\label{mathematicalexpectation1}
{V_{{\pi ^ * }}}\left( s \right) = \mathop {\max }\limits_{a \in A} \left[ {r\left( {s,a} \right) + \gamma \sum\limits_{{s^ * } \in S} {{P_{s,{s^ * }}}{V_{{\pi ^ * }}}\left( {{s^ * }} \right)} } \right].
\end{equation}
where Bellman's optimality criterion states that there is one optimal strategy in a single environment setting. With the above circumstances, we define the states and the actions of Q-learning in our scenario as below

\begin{enumerate}
\item \textbf{States space:} states are matrices with size of $S \times M$, and the total number of states is ${F^{S \times M}}$.
\item \textbf{Actions space:} actions we take here is to change the state, and thus the actions are increase one or decrease one of the element state matrix. As a result, the total number of actions are $2 \times S \times M + 1$, with the last action ${\left[ 0 \right]_{S \times M}}$, which means that the state does not need to change.
\item \textbf{Reward function:} the reward that we give here is based on objective function MOS:

\begin{equation}\label{rewardmos}
r\left( {{a_t},{s_t},{s_{t + 1}}} \right) = \left\{ {\begin{array}{*{20}{c}}
0&{{\rm{MO}}{{\rm{S}}_{{s_{t + 1}}}} \ge {\rm{MO}}{{\rm{S}}_{{s_t}}}}\\
1&{{\rm{MO}}{{\rm{S}}_{{s_{t + 1}}}} < {\rm{MO}}{{\rm{S}}_{{s_t}}}.}
\end{array}} \right.
\end{equation}

%\item Content placement: The optimal caching, i.e., the optimal content placement $s*$, is the one with the maximum $Q$ values
%
%\begin{equation}\label{optimalQ}
%{s^ * } = \mathop {\max }\limits_{a \in A} \left\{ {Q\left( {s,a} \right)} \right\}.
%\end{equation}

\item For Q value (or state-action value) update, Bellman Equation is applied,

\begin{equation}\label{qupdate}
Q\left( {s,a} \right) \leftarrow \left( {1 - \alpha } \right)Q\left( {s,a} \right) + \alpha \left[ {r + \gamma {{\max }_{a'}}Q\left( {s',a'} \right)} \right].
\end{equation}
where $\alpha  \in \left( {0,1} \right)$ denotes the learning rate, which determines the speed of learning, (i.e., a larger $\alpha$ value leads to a faster learning process, yet may result in non-convergence of learning process, and a smaller $\alpha$ value leads to a slower learning process). $\gamma  \in \left( {0,1} \right)$ denotes the discount factor, which determines the balance between historical Q value and future Q value, (i.e., a larger $\gamma$ value means that the future Q value is more important, and a smaller $\gamma$ value will balance more on the current Q value).

\end{enumerate}

\begin{figure} [t!]
 \centering
 \includegraphics[width=3.5in]{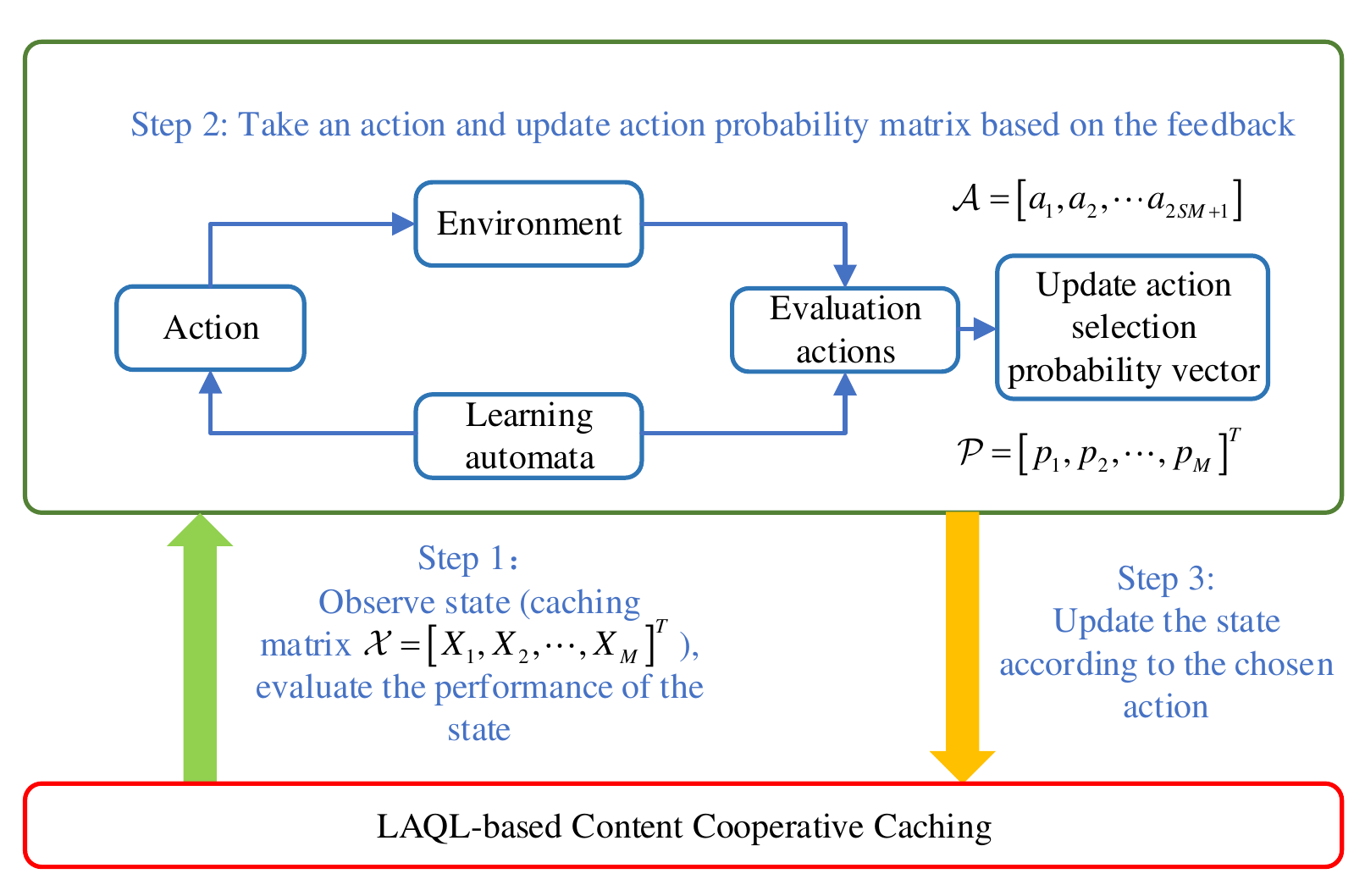}
 \centering
 \caption{An illustration of LAQL-based content cooperative caching.}\label{Qlearningflow}
\end{figure}

The details of Q-learning based wireless content caching algorithm are illustrated in \textbf{Algorithm}~\textbf{\ref{Qlearning}} and \textbf{Algorithm}~\textbf{\ref{Qlearningtest}}. The state matrix $\left( {{s_t},{a_t},{s_{t + 1}},Q\left( {s,a} \right)} \right)$ is stored in a matrix to train the Q matrix. Using greedy Q-learning algorithm to search best state-action policy may cause an issue that some states may never be visited. For this reason, the agent might become stuck at certain states without even knowing about better possibilities. Therefore we propose a LAQL based cooperative caching algorithm. LAQL based cooperative caching algorithm is a tradeoff between exploration and exploitation, where exploration means to explore the effect of unknown actions, while exploitation means to remain at the current optimal action since learning.

Q-learning achieves optimal solution if $\varepsilon  = 1$. However, in this circumstance, the probability that Q-learning achieves local-optimal solution is much higher than that Q-learning achieves optimal solution. For this reason, we normally set $\varepsilon  < 1$ to explore all states, and to obtain quasi-optimal solution.

%\begin{remark}\label{remark:Qlearningandexhaustivesearch2}
%The stability of Q-learning based content cooperative caching is achieved only when both the policy and the value function are optimal. As an off-policy algorithm, Q-learning is capable of learning from existing policy exploration.
%\end{remark}
%A learning automata (LA) is an adaptive decision-making unit which learns the optimal action from a set of actions offered by the environment it operates in. At each iteration, the LA selects one action, which triggers either a stochastic reward or a penalty as a response from the environment. Based on the response and the knowledge acquired in the past iterations, the LA adjusts its action selection strategy in order to make a ``wiser'' decision in the next iteration. In such a way, the LA, even though it lacks a complete knowledge about the Environment, is able to learn through repeated interactions with the Environment, and adapts itself to the optimal decision.
\subsection{Learning Automata Based Q-learning For Cooperative Caching}

\begin{algorithm}[h]
\caption{LA Based Q-learning Algorithm for Content Cooperative Caching (Training Stage)}
\label{Qlearning}
\begin{algorithmic}[1]
\REQUIRE Caching matrix of state $s$, caching action $a$, and $Q\left( {s,a} \right)$.

    \STATE Initialise Q-table $Q\left( {s,a} \right)$, state $s$, and action selection probability vector in LA for each state $s$.\

    \REPEAT
    \FOR{each episode}
    \FOR{each step}
    \STATE Choose cooperative caching action $a$ according to \textbf{Learning Automata Based Action Selection Scheme:}\\
      %\begin{algsubstates}
        5.1: The agent chooses one of the actions according to the distribution of ${P_a}(t)$ in the LA for the current state;\\
        5.2: Update the reward probability estimation based on the feedback from the environment for the chosen action.\\
        5.3: \textbf{If} action is active, i.e., a reward is achieved. \textbf{Then}
        \begin{equation}\label{autoqu4}
        \left\{ {\begin{array}{*{20}{l}}
        {{p_i}\left( {t + 1} \right) = max\left\{ {{p_i}\left( t \right) - \Delta ,0} \right\},i \ne h}\\
        {{p_h}\left( {t + 1} \right) = 1 - \sum\limits_{i,i \ne h} {{p_i}\left( {t + 1} \right)} }
        \end{array}} \right.
        \end{equation}
        \textbf{Else}
        $P_a^{}\left( {t + 1} \right) = P_a^{}\left( t \right).$

        \textbf{EndIf}
      %\end{algsubstates}
    \STATE Take the chosen action $a$, thereafter calculate reward $r$ of the action $a$ and state $s$;\
    \STATE Update Q-table: calculate Q value using the defined reward function in (\ref{qupdate});\
    \STATE State update: $s \leftarrow s'$;\
    \ENDFOR
    \ENDFOR
    \UNTIL {state $s$ does not change.}
    \ENSURE Q-table $Q\left( {s,a} \right)$, and content cooperative caching matrix of BSs $X$.

\end{algorithmic}
\end{algorithm}

An LA is an adaptive decision-making unit which learns the optimal action from a set of actions offered by the environment it operates in. In our scenario, the optimization problem has complex structures and long horizons, which poses a challenge for the reinforcement learning agent due to the difficulty in specifying the problem in terms of reward functions as well as large variances in the learning agent. We address these problems by combining LA with reinforcement learning for cooperative caching.

The goal of Q-learning is to find an optimal policy that maximize the expected sum of discounted return:

\begin{equation}\label{goalofFSA}
{\pi ^*} = \mathop {\arg \max {E^\pi }}\limits_\pi  \left[ {\sum\limits_{t = 0}^{T - 1} {{\gamma ^{t + 1}}\widetilde r\left( {{{\widetilde s}_t},{{\widetilde s}_{t + 1}}} \right)} } \right].
\end{equation}

In our scenario, the state space and action space are discrete. Different from traditional Q-learning, the agent of LAQL algorithm chooses the actions according to the effectiveness of actions (i.e., whether the action is a success or failure). The LAQL algorithm consists of three steps. In each iteration, firstly, an action probability vector is maintained: ${P_a}\left( t \right) = \left[ {{p_1}\left( t \right),{p_2}\left( t \right), \cdots ,{p_{{n_a}}}\left( t \right)} \right]$, where ${{n_a}}$ represents the number of actions of the target state and $\sum\limits_{i = 1 \cdots {n_a}} {{p_i}} \left( t \right) = 1$. We select an action according to the probability distribution ${P_a}\left( t \right)$. Secondly, based on the current feedback, the maximum likelihood reward probability is estimated for the chosen action in the current iteration. Suppose the $i$th action of the target state is chosen in the current iteration, then the reward probability ${\widetilde d_i}\left( t \right)$ is updated as~\cite{Zhang2016,Lei2019JNNLS}:

\begin{equation}\label{autoqu1}
\begin{array}{l}
{u_i}\left( t \right) = {u_i}\left( {t - 1} \right) + \left( {1 - r\left( t \right)} \right)\\
{v_i}\left( t \right) = {v_i}\left( {t - 1} \right) + 1\\
{\widetilde d_i}\left( t \right) = {\raise0.7ex\hbox{${{u_i}\left( t \right)}$} \!\mathord{\left/
 {\vphantom {{{u_i}\left( t \right)} {{v_i}\left( t \right)}}}\right.\kern-\nulldelimiterspace}
\!\lower0.7ex\hbox{${{v_i}\left( t \right)}$}},
\end{array}
\end{equation}
where ${u_i}\left( t \right)$ denotes the number of times that action $i$ has been rewarded when selected. ${v_i}\left( t \right)$ represents the number of times that action $i$ has been selected.
%
%When we get a positive reward value from an action $i$ at time $t$:
%
%\begin{equation}\label{autoqu2}
%\left\{ \begin{array}{l}
%{p_i}\left( {t + 1} \right) = {p_i}\left( t \right) + \alpha \left( {1 - {p_i}\left( t \right)} \right)\\
%{p_j}\left( {t + 1} \right) = \left( {1 - \alpha } \right){p_j}\left( t \right)
%\end{array} \right.
%\end{equation}
%where $j$ denotes the other actions apart from action $i$. $\alpha$ denotes the reward factor.

\begin{algorithm}[h]
\caption{LA Based Q-learning Algorithm for Content Cooperative Caching (Testing Stage)}
\label{Qlearningtest}
\begin{algorithmic}

\REQUIRE state $s$, action $a$.
    \STATE Generate a content cooperative caching matrix of network randomly, and set it as the initiating state $s$;\
    \FOR{each iteration}
    \STATE Choose an action of the current state according to \textbf{Learning Automata based Action Selection Scheme};
    \STATE Calculate reward, and update state;
    \ENDFOR

\ENSURE Content cooperative caching matrix of BSs $X$.

\end{algorithmic}
\end{algorithm}

Thirdly, if $r\left( t \right) = 0$, meaning a reward is achieved, the LAQL agent increases the probability of selecting the current best action based on the current reward estimates of all actions. In more details, if ${\widetilde d_h}\left( t \right)$ is the largest element among all reward estimates of all actions, we update ${p_i}\left( t \right),i \in \left\{ {1,2, \cdots ,{n_a}} \right\}$ as:

\begin{equation}\label{autoqu3}
\left\{ {\begin{array}{*{20}{l}}
{{p_i}\left( {t + 1} \right) = max\left\{ {{p_i}\left( t \right) - \Delta ,0} \right\},i \ne h}\\
{{p_h}\left( {t + 1} \right) = 1 - \sum\limits_{i,i \ne h} {{p_i}\left( {t + 1} \right)} }
\end{array}} \right.
\end{equation}
where $\Delta  = {\raise0.7ex\hbox{$1$} \!\mathord{\left/
 {\vphantom {1 {{n_a}\kappa }}}\right.\kern-\nulldelimiterspace}
\!\lower0.7ex\hbox{${{n_a}\kappa }$}}$, and $\kappa $ being a positive integer.

If $r\left( t \right) = 1$, then $P_a^{}\left( {t + 1} \right) = P_a^{}\left( t \right)$.

%\title{Lemma for LA} % insert title - use \\ if it requires more than one line.
Clearly, although the converged $Q(\overrightarrow s, \overrightarrow a)$ value indicates the long term average reward of the actions for a given state $\overrightarrow s $ and action $\overrightarrow a$, due to the random nature of the studied scenario, i.e., the uncertainty of the destination state upon an action, the optimal action may not surely return a higher reward than the suboptimal actions at a certain time instant when we select it. Therefore, although it is important to apply LA to balance exploration and exploitation, it is necessary to show that LA can converge to the optimal action with high probability in this random environment.

\begin{lemma}\label{lemma:exclude_lieral}
Given that $Q(s,a)$ is capable of converging to the optimal Q value, the LA corresponding to every state $s$ can select the optimal action with arbitrary high probability.

\begin{IEEEproof}
See Appendix B~.
\end{IEEEproof}
\end{lemma}

\begin{remark}\label{remark:AQLoptimal}

The proposed LA based action selection scheme enables the states in Q-table to select the optimal action.

\end{remark}

%\subsection{Computational Issues}
%
%In the process of cooperative caching between different BSs, the complexity of Q-learning based caching depends primarily on the number of BSs $M$, caching capacity of BS $S$, the number of contents $F$, as can be seen in TABLE~\ref{table3}. We consider an example where $F=10$, $S=4$, $M=10$, $K=3$, the computational complexity is shown in the second column in TABLE~\ref{table3}. As can be seen from TABLE~\ref{table3}, Q-learning based caching takes 4000 operations, whereas optimal caching takes ${10^{4 \times 10}}$.
%
%\begin{table}[h]
%	\caption{The Number of Operations Required By Different Schemes}
%	\centering
%	\begin{tabular}{|c|c|c|}\hline
%		Method & Number of Operations & NO. of operations\\\hline
%		Optimal caching & ${F^{SM}}$ & ${10^{4 \times 10}}$  \\\hline
%        K-means Algorithm & ${F^{SK}}$ & ${10^{4 \times 3}}$ \\\hline
%        Q learning Algorithm & ${FSM}$ & $10 \times 4 \times 10$ \\\hline
%	\end{tabular}
%\label{table3}
%\end{table}
%
%In terms of storage requirements, however, the optimal caching method possesses lowest number of memory unites since it does not need to memorize much knowledge. The table of Q learning requires a higher number of memory units, up to the maximum as ${F^{SM}} \times \left( {2 \times S \times M + 1} \right) $. Nevertheless, it should be mentioned that it is possible to reduce the storage requirements of Q table by using neural network (i.e., Deep Q Networks).

\subsection{Computational Issues}

The computational complexities of traditional Q-learning and LAQL are quite different. In the process of content caching in different BSs, the complexity of Q-learning based cooperative caching depends primarily on the number of BSs $M$, caching capacity of BS $S$ the number of contents $F$, as can be seen in TABLE~\ref{table3}. We consider an example where $F=10$, $S=4$, $M=10$, the computational complexity is shown in the second column in TABLE~\ref{table3}. As can be seen from TABLE~\ref{table3}, LAQL based cooperative caching takes 32000 operations, whereas optimal caching takes ${10^{4 \times 10}}$ and $\varepsilon $-greedy Q-learning takes 4000.

\begin{table}[h]
                \caption{Number of Operations Required For Different Schemes}
                \centering
                \begin{tabular}{|c|c|c|}\hline
            Method & Number of Operations & Numerical Example\\\hline
            Optimal caching & ${F^{SM}}$ & ${10^{4 \times 10}}$  \\\hline
            $\varepsilon $-greedy Q-learning Algorithm & ${FSM}$ & $10 \times 4 \times 10$ \\\hline
            Learning automata based Q-learning Algorithm & $FSM \times 2SM$ & $10 \times 4 \times 10 \times 2 \times 4 \times 10 $ \\\hline
                \end{tabular}
\label{table3}
\end{table}

In terms of storage requirements, however, the optimal caching method possesses lowest number of memory unites since it does not need to memorize much knowledge. The table of LAQL and $\varepsilon $-greedy Q-learning requires a higher number of memory units, the maximum of which is ${F^{SM}} \times \left( {2 \times S \times M + 1} \right) $.

\section{Numerical Results}\label{section:Simulationresults}

In this section, simulations are conducted to evaluate the performance of the proposed algorithms. We first consider the situation that the users move in a random walk model, after that we utilize user mobility data collected from smartphones to test the performance of neural network based user mobility prediction. The position data of on single user is collected from daily life using GPS-tracker app on an Android system smartphone. We open the APP when we start the journey, and record the positions during the journey. Meanwhile, we also consider random walk model for content popularity. We use random walk model to test the prediction accuracy of our proposed neural network for content popularity prediction. For the simulation setup, we consider a cooperative square region with length of a side 4 km. There are $M=2$ BSs in the region, and $N=100$ users are randomly and independently distributed in the region. We assume that the total number of contents is $F=10$. All the simulations are performed on a desktop with an Intel Core i7-7700 3.6 GHz CPU and 16 GB memory.

\begin{table}[h]
\caption{Parameter Configurations}
	\centering
	\begin{tabular}{|c|c|c|}\hline
		Parameter&Description&Value\\\hline
		$F$ & Content library size & 10  \\\hline
        $M$ & Number of BSs & 2  \\\hline
        $N_u$ & Number of Users & 100  \\\hline
        $L$ & Size of content & 1  \\\hline
        %$f_c$ & Carrier frequency & 1.8 GHz  \\\hline
%        $d_0$ & free-space reference distance & 5 m\cite{Chen2017TWC}  \\\hline
        $B$ & Bandwidth & 20MHz\cite{Chen2017TWC}  \\\hline
        ${{P_t}}$ & Transmit power of each BS & 20 dBm\cite{Chen2017TWC}  \\\hline
        $\alpha$ & Path loss exponent & 3  \\\hline
        $S$ & Caching capacity & 4  \\\hline
        ${\sigma ^2}$ & Gaussian noise power & -95 dBm\cite{Chen2017TWC}  \\\hline
        ${C_1}$ & constant & 1.120\cite{Karedal2008TWC}  \\\hline
        ${C_2}$ & constant& 4.6746\cite{Karedal2008TWC}  \\\hline

	\end{tabular}

	\label{table2}
\end{table}

To investigate the performance of the proposed algorithms, three different algorithms are simulated for wireless cooperative caching system, respectively. Firstly, we consider the global optimal content caching scheme called ``Optimal Caching''. In ``Optimal Caching'', an exhaust search is exploited over all combinations of content index and BS index to find optimal content caching matrix. The benchmark we used is non-cooperative caching, i,e,. the BSs do not cooperative with each other. In other words, the BSs cache the most popularity contents independently. Another benchmark is random caching, i.e., the BSs do not know the prior information such as content popularity and user mobility, hence the BSs cache the contents randomly.

%\begin{figure*}
%
%    \centering
%
%    \subfigure[Mean Squared Error (mse) of SFBC-NN for content popularity prediction.]{
%    \begin{minipage}{7cm}
%    \centering
%    \includegraphics[width=7cm]{eps/popumse.eps}\label{popu1}
%    \end{minipage}
%    }
%    \subfigure[Performance of SFBC-NN for content popularity prediction.]{
%    \begin{minipage}{7cm}
%    \centering
%    \includegraphics[width=7cm]{eps/popuperformance.eps}\label{popu2}
%    \end{minipage}
%    }
%    \caption{The SFBC-NN based content popularity prediction.}\label{popuresult}
%\end{figure*}

\begin{figure} [t!]
\centering
\includegraphics[width=3.5in]{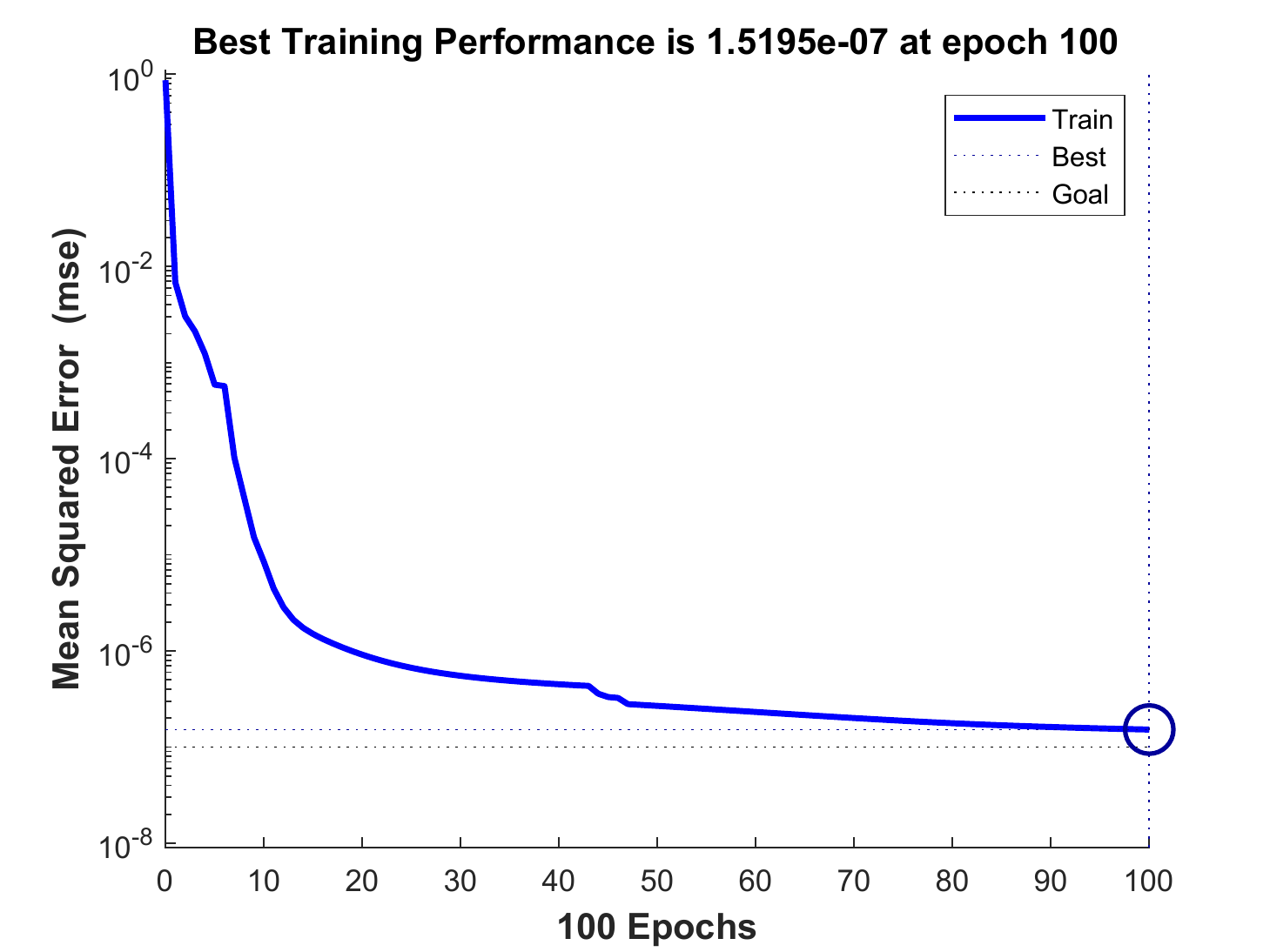}
 \caption{Mean Squared Error (MSE) of SFBC-NN for content popularity prediction.}\label{popu1}
\end{figure}

\begin{figure*}

    \centering
    \subfigure[Comparisons of real trajectory and prediction trajectory.]{
    \begin{minipage}{5cm}
    \centering
    \includegraphics[width=5cm]{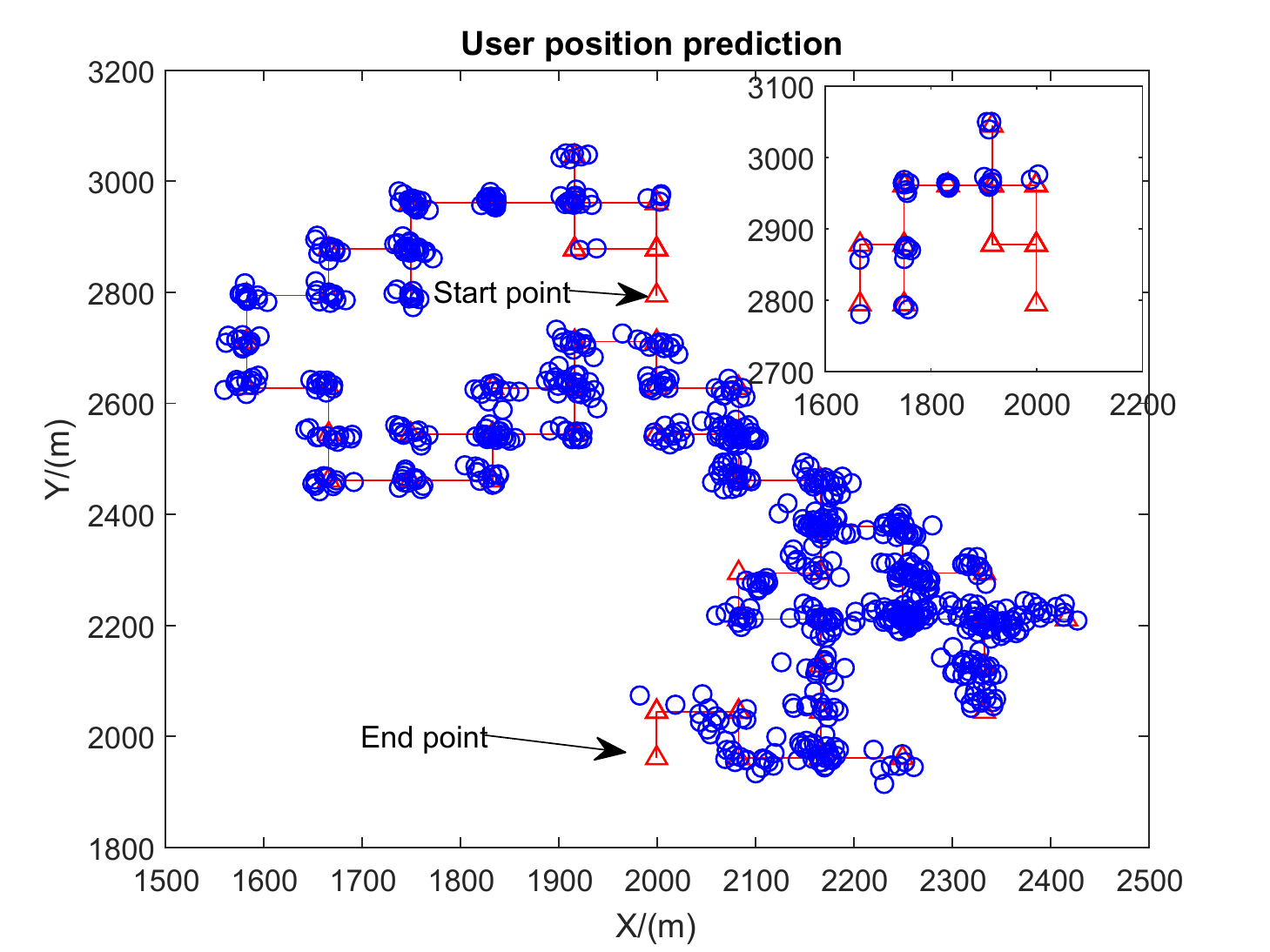}\label{Userpositionprediction}
    \end{minipage}
    }
    \subfigure[Comparisons of real trajectory and predicted trajectory in X and Y axis]{
    \begin{minipage}{5cm}
    \centering
    \includegraphics[width=5cm]{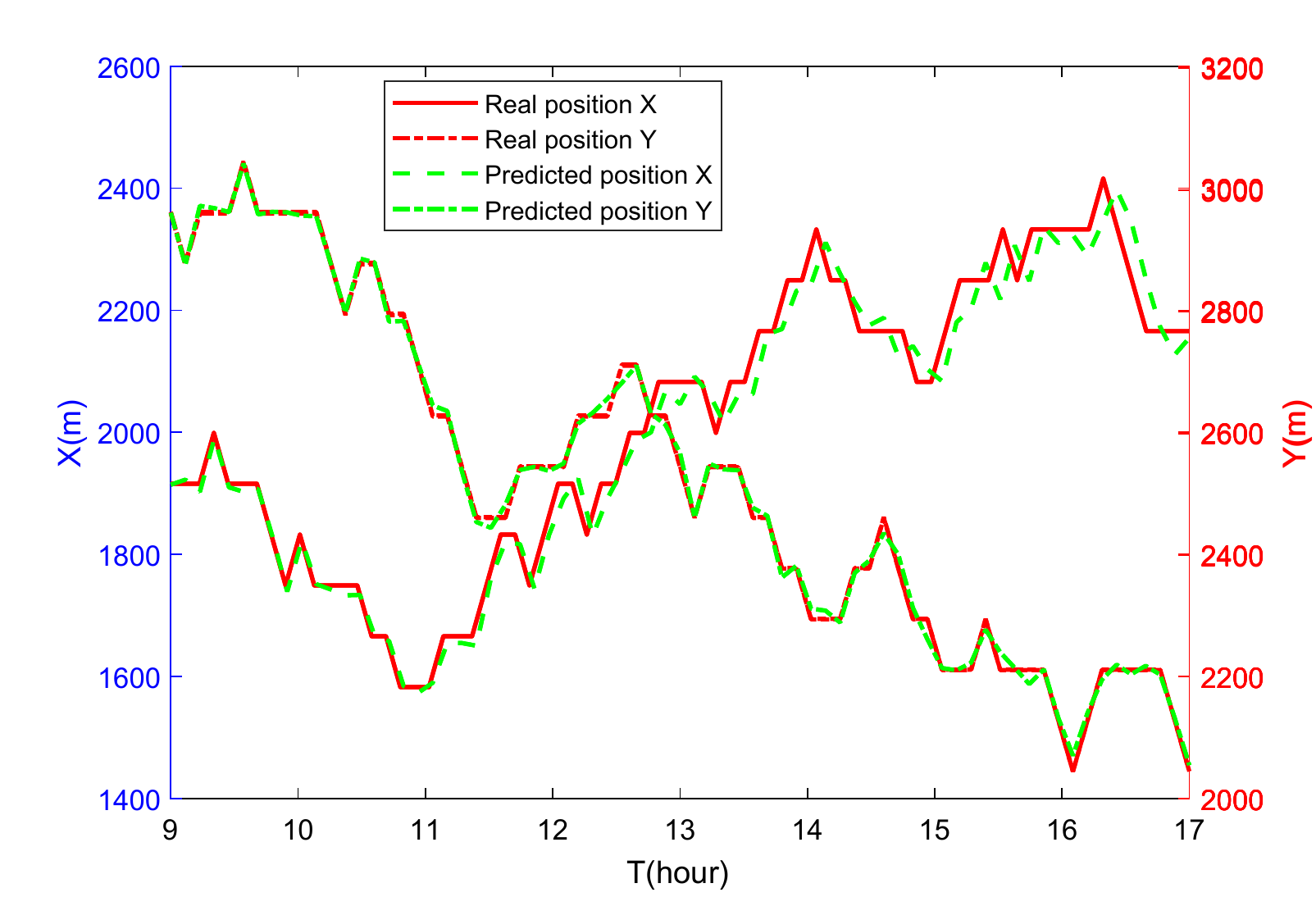}\label{XYprediction}
    \end{minipage}
    }
    \subfigure[Performance with the Number of Nodes and the Number of iterations.]{
    \begin{minipage}{5cm}
    \centering
    \includegraphics[width=5cm]{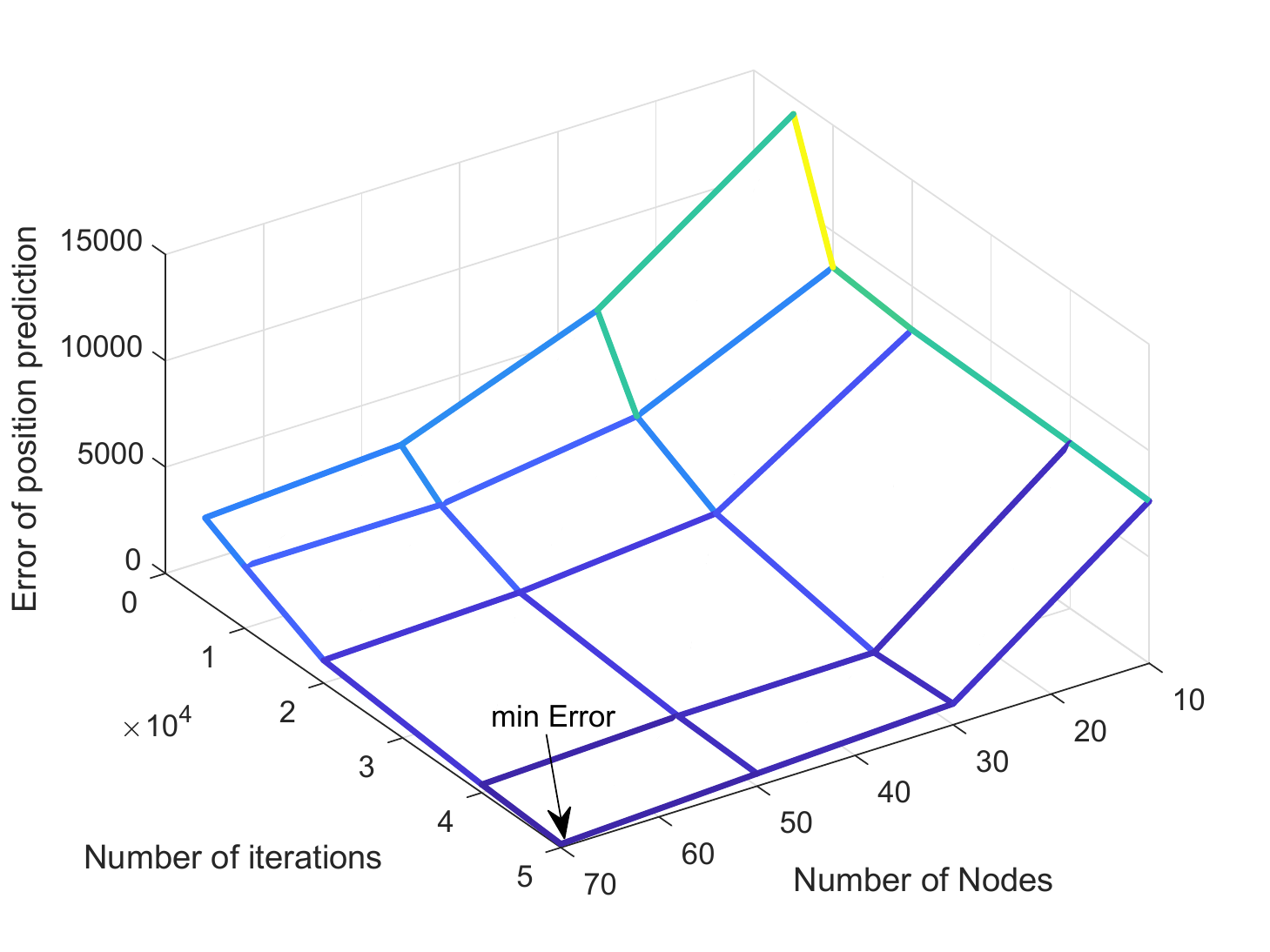}\label{errorpositionprediction}
    \end{minipage}
    }
    \caption{The SFBC-NN prediction of users mobility (data from simulation).}\label{SFBC-NNprediction}
\end{figure*}

\begin{figure*}

    \centering
    \subfigure[Trajectory of a user in Google map.]{
    \begin{minipage}{5cm}
    \centering
    \includegraphics[height=1.8in]{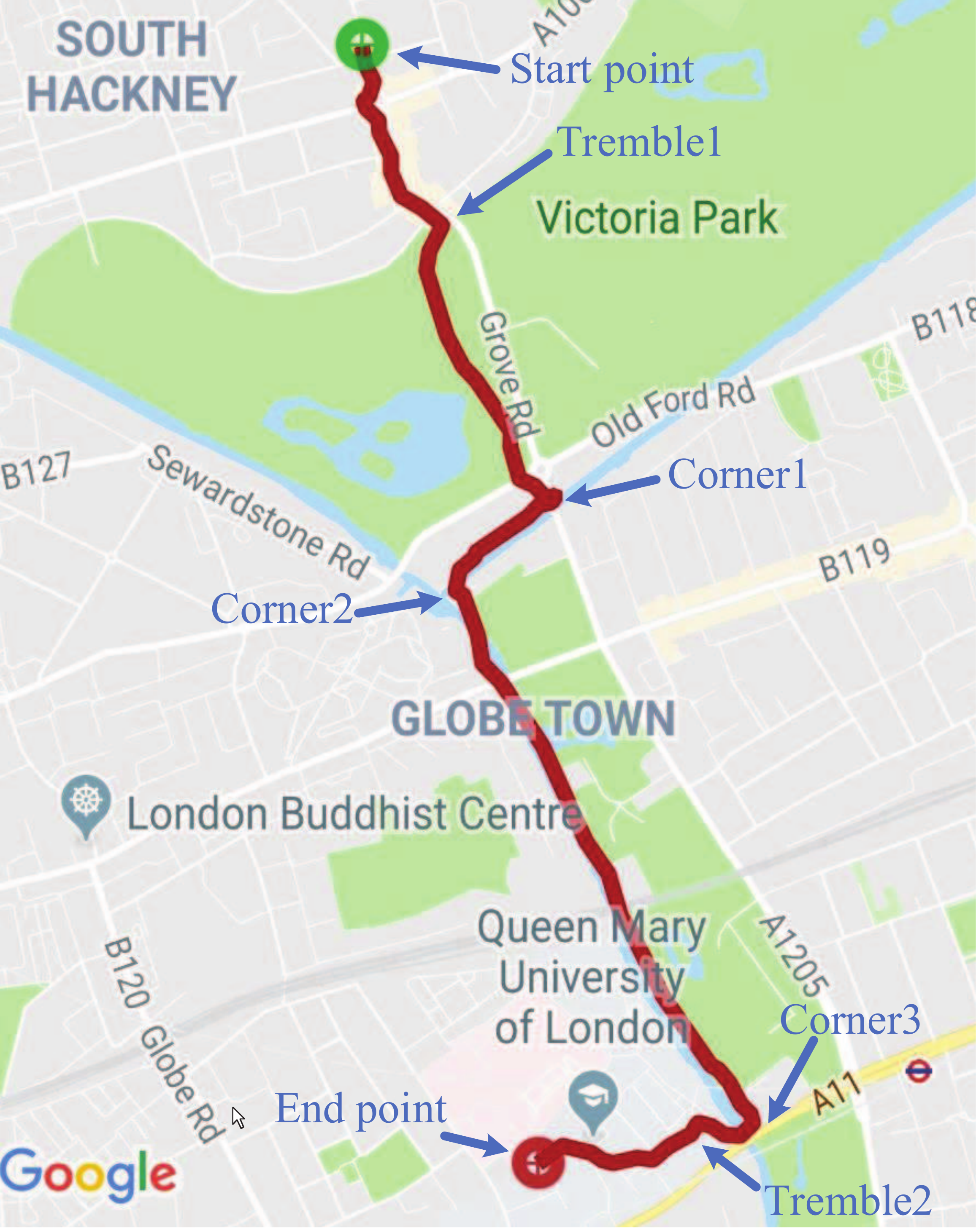}\label{realdata1}
    \end{minipage}
    }
    \subfigure[Comparisons of real trajectory and prediction trajectory.]{
    \begin{minipage}{5cm}
    \centering
    \includegraphics[width=5cm]{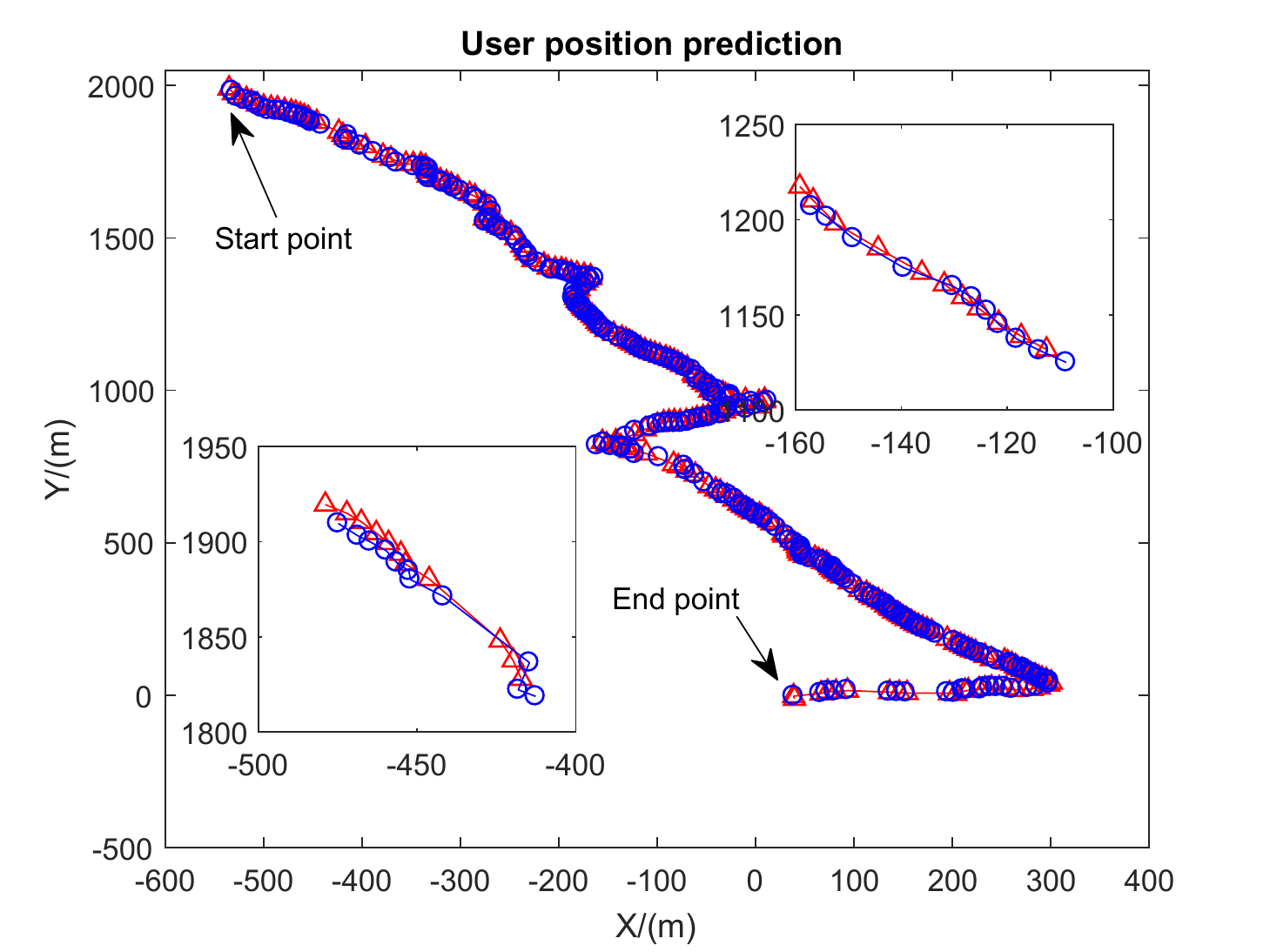}\label{realdata2}
    \end{minipage}
    }
    \subfigure[Performance of SFBC-NN prediction (data measured from smartphone).]{
    \begin{minipage}{5cm}
    \centering
    \includegraphics[width=5cm]{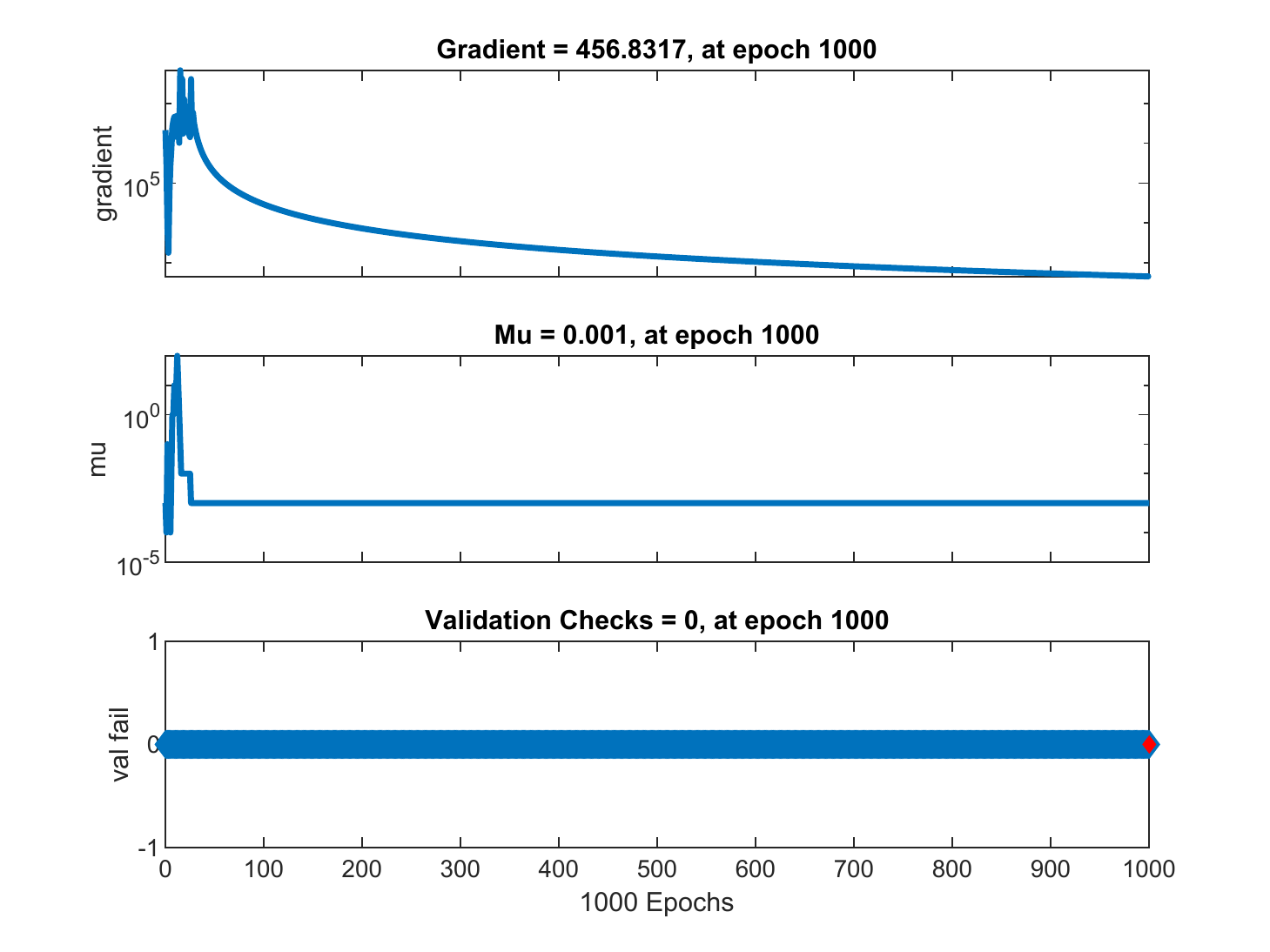}\label{realdata3}
    \end{minipage}
    }
    \caption{The SFBC-NN prediction of user's mobility (data from smartphone).}\label{realdata}
\end{figure*}

\subsection{Simulation Results of Content Popularity Prediction}

In this subsection, we utilize history data of content popularity to predict content popularity in the future. More precisely, content popularity in 5 time slots are provided as SFBC-NN input, while the output is content popularity in the next slot. As is shown in Fig. \ref{popu1}, MSE of the network decreases with epochs. It reached bottom after 50 epochs.

\subsection{Simulation Results of User Mobility Prediction}

%In order to analysis the mobility of users, we assume that users' positions change periodically. Firstly, we generate user's positions through random work model. We randomly produce the users' positions every 5 minutes, and then use one user's one hour positions (i.e., 12 samples), to predict the user's position in the next hour. Fig. \ref{SFBC-NNprediction} evaluates the prediction accuracy of SFBC-NN based user position prediction. In Fig.~\ref{Userpositionprediction}, the red line shows user real trajectory, and blue points are predictions of user positions. As is shown is the Fig.~\ref{Userpositionprediction} and~\ref{XYprediction}, the predicted trajectory converges to the real trajectory, which clarifies that the trained weight parameters of SFBC neural network is capable of predict user position.

In order to analysis the mobility of users, firstly, we generate user's positions through random work model. Fig. \ref{SFBC-NNprediction} evaluates the prediction accuracy of SFBC-NN based user position prediction. In Fig.~\ref{Userpositionprediction}, the red line shows user real trajectory, and blue points are predictions of user positions. As is shown is the Fig.~\ref{Userpositionprediction} and~\ref{XYprediction}, the predicted trajectory converges to the real trajectory, which clarifies that the trained weight parameters of SFBC neural network is capable of predict user position.

Fig.~\ref{errorpositionprediction} shows the error of position prediction, where the error decreases with the number of nodes in SFBC-NN and that of iterations. This is because the increased number of nodes and iterations enhances the fitting capability of the network, thus strengthen the prediction accuracy. Due to the fact that a user position has only two dimensions, the simulation results show that SFBC-NN is powerful to deal with user mobility prediction.
%However due to the fact that the user mobility time interval we input here is just one day, the network we need here is simple, and the number of nodes and that of iterations are also small. Due to the fact that a user position has only two dimensions, the simulation results show that SFBC-NN is powerful to deal with user mobility prediction.

Sociological researches have provided us various methods to collect people's moving trajectory. However, it consumes a lot of resources (i.e., time, manpower, and money) to collect a large number of data. For this reason, we collect movements of one user, to validate the accuracy of our proposed prediction algorithm. In Fig.~\ref{realdata1}, we use measured data from smartphone to test the prediction accuracy of the SFBC-NN. The data is collected from daily life using GPS-tracker app on an Android smartphone. Fig.~\ref{realdata1} shows the trajectory of user in Google Maps, and there are trembles and corners in the trajectory, which will effect prediction accuracy. The results indicated in Fig.~\ref{realdata2} confirm the accuracy of SFBC neural network. This phenomenon is also confirmed by the insights in \textbf{Remark 2}. Fig.~\ref{realdata3} shows performance of SFBC-NN based user mobility prediction. By indicating the gradual decreasement of gradient with training epoch number. Similarly mu also decreases to a stable state at approximately 20 epoches. And there are no fail observed.

\begin{figure*}

    \centering

    \subfigure[Q learning based caching.]{
    \begin{minipage}{7cm}
    \centering
    \includegraphics[width=7cm]{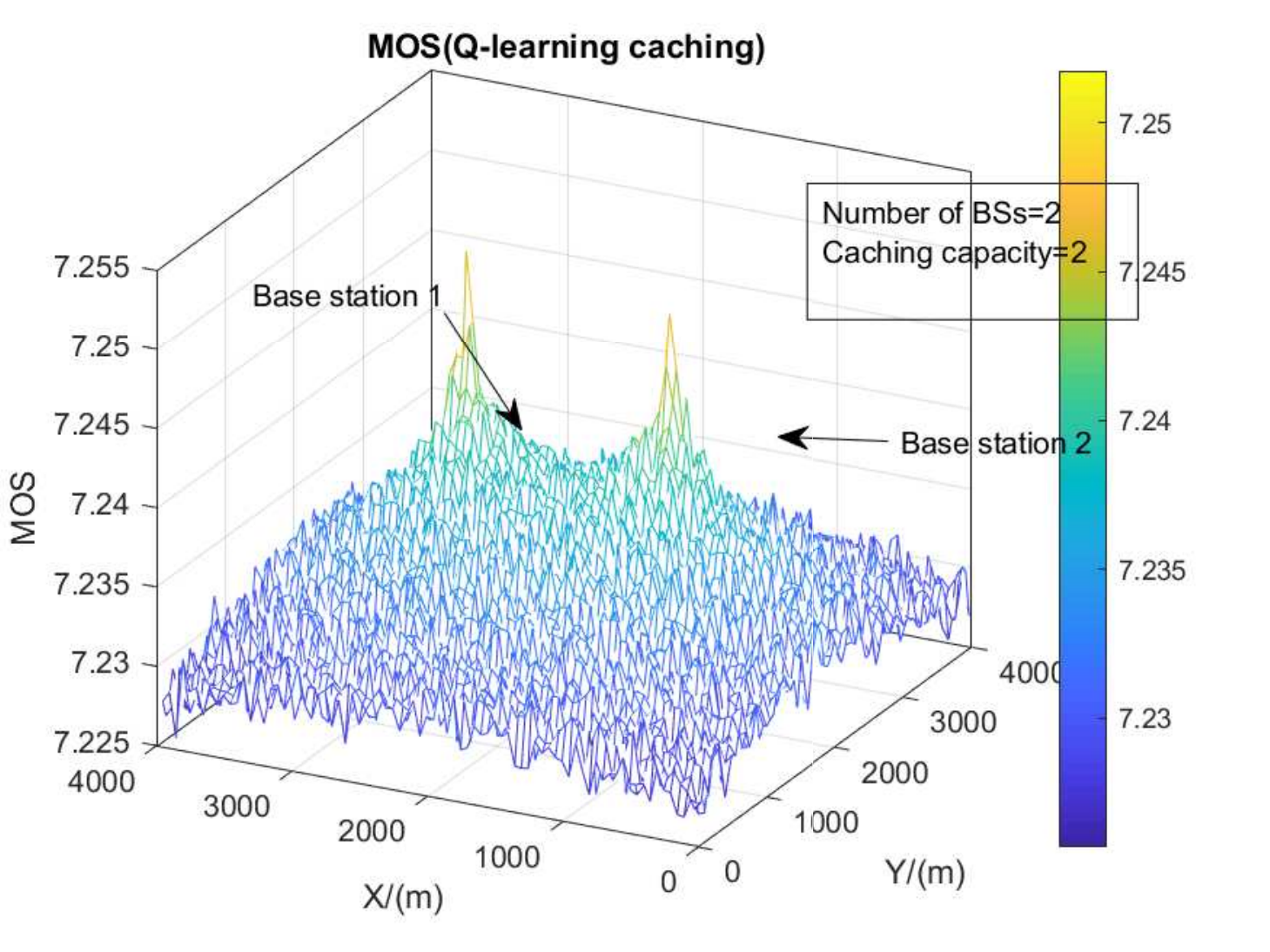}
    \end{minipage}
    }
    \subfigure[Optimal caching.]{
    \begin{minipage}{7cm}
    \centering
    \includegraphics[width=7cm]{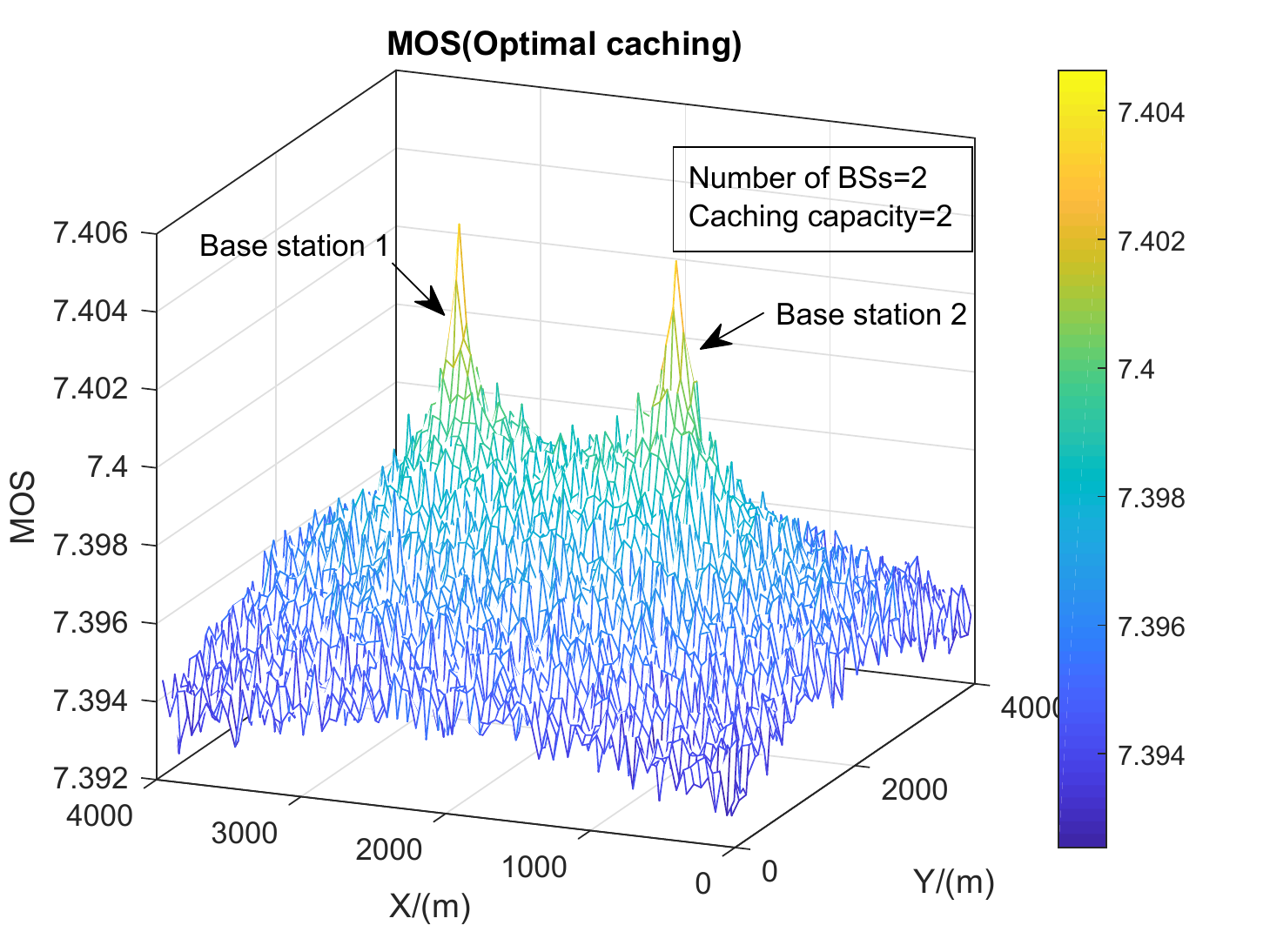}
    \end{minipage}
    }
    \caption{The MOS of users distributes in different positions of the area for different caching methods.}\label{QlearningcachingM2cc3}
\end{figure*}

\begin{figure} [t!]
\centering
\includegraphics[width=3.5in]{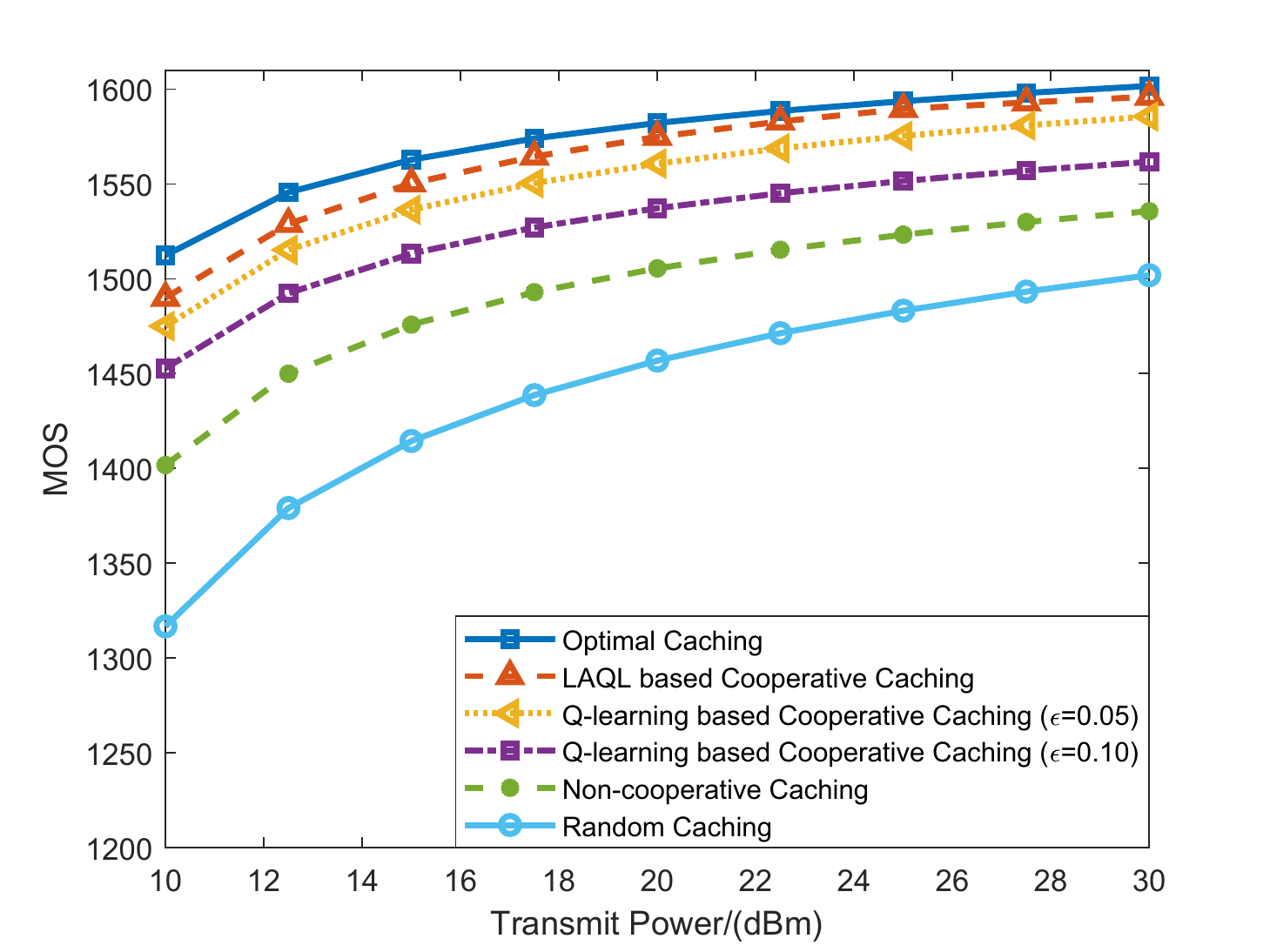}
 \caption{The MOS of different cache scheme vs. transmit power and the number of BSs and users.}\label{Qlearningcaching}
\end{figure}

%\begin{figure} [t!]
%\centering
%\includegraphics[width=4in]{eps/cachingmobility1.eps}
% \caption{The overall MOS of users as a function of time period for different caching algorithms. }\label{cachingwithmobility}
%\end{figure}

\begin{figure} [t!]
\centering
\includegraphics[width=3.5in]{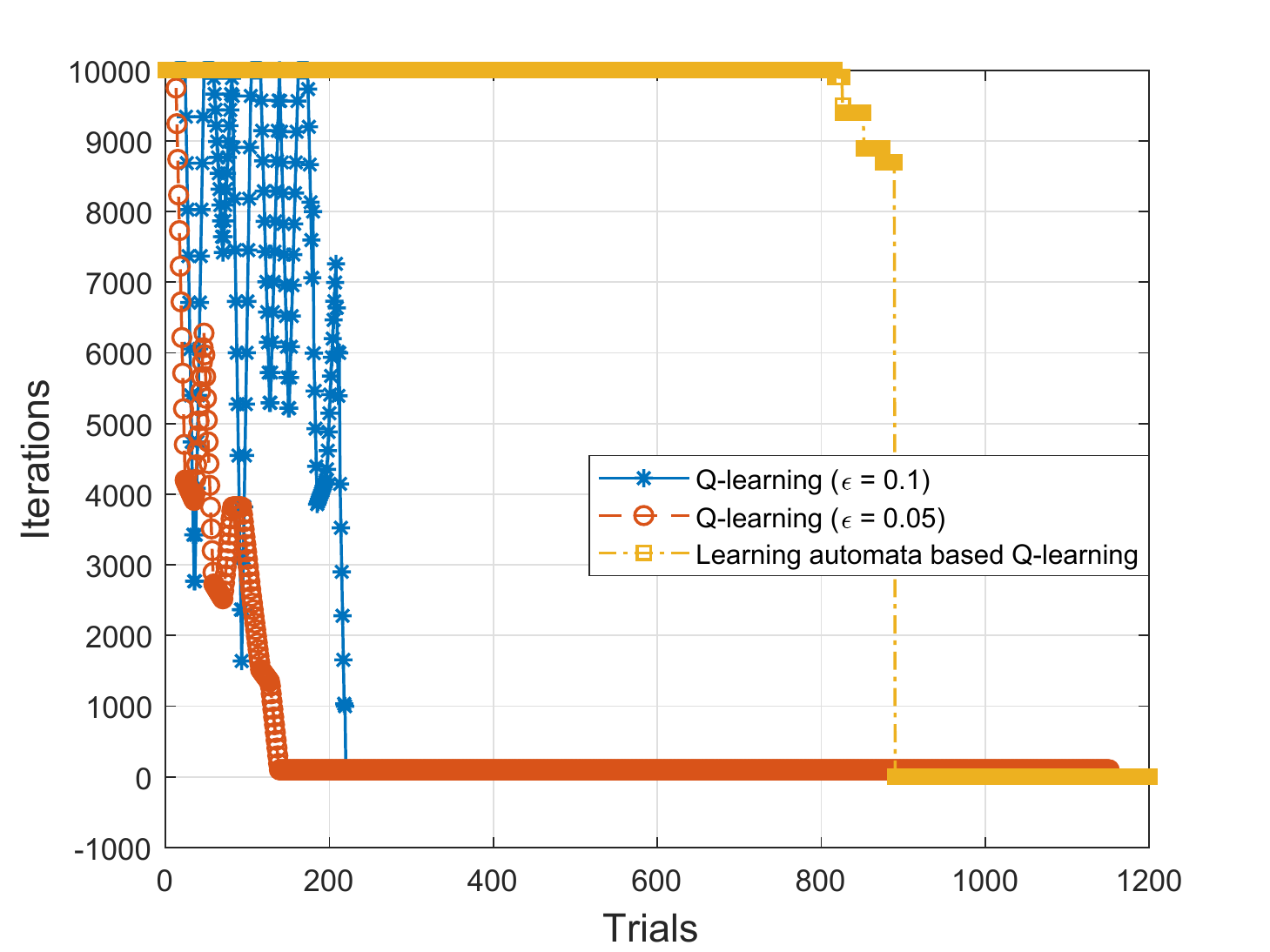}
 \caption{The reward curves of learning automata based Q-learning and conventional Q-learning over trials. }\label{automatagididedqlearning}
\end{figure}

\subsection{Simulation Results of LAQL Based Cooperative Caching}

In this subsection, we compare the proposed LAQL based cooperative caching with the above mentioned traditional caching schemes. Fig.~\ref{QlearningcachingM2cc3} shows QoE distribution of users in this rectangle area, with different caching schemes, (i.e., Q-learning based caching and optimal caching). It can be observed that users near BSs have better QoE than those who are far from BSs. This is easy to understand because the distance between a user and a BS has important influence on MOS of the user, and also this can be further explored to analysis the deployment of BSs (i.e., deployment schemes and density of BSs).

Fig.~\ref{Qlearningcaching} shows the overall MOS of users as a function of transmit power. We first adjust the learning rate $\alpha$ and discount factor $\gamma$ of traditional Q-learning algorithm, because $\alpha$ and $\gamma$ have significant influence of Q-learning. After several adjustment, we set $\alpha=0.75$ and $\gamma=0.6$. After that, we set the learning rate and discount factor of LAQL is the same as traditional Q-learning, aiming at reducing the interference of learning rate and discount factor between traditional Q-learning and LAQL. As can be seen from this figure, MOS increase with greater transmit power. We see that the performance of LAQL based cooperative caching is closer to optimal solution than traditional Q-learning, non-cooperative caching and random caching. This is because the action selection scheme of LAQL enables the agent to choose optimal action in each state, while action selection scheme in traditional Q-learning is based on a stochastic mechanism (i.e., $\epsilon$-greedy). Fig.~\ref{Qlearningcaching} also shows that cooperative caching outperforms random caching and non-cooperative caching. The reason is that the cooperative caching takes user mobility and content popularity into consideration while random caching does not. Meanwhile, the cooperative caching is capable of better balance the caching resource in different BSs than non-cooperative caching.
%Fig.~\ref{cachingwithmobility} shows the overall MOS of users in the network for different caching algorithms. To reduce computational complexity, we set the number of users as one hundred, the total number of contents as ten and the caching capacity of each BS as two. The simulation result is shown in Fig.~\ref{cachingwithmobility}. The performance of different algorithms decreases with time. The optimal caching outperforms Q-learning based caching, non-cooperative caching and random caching. Additionally, as can been seen from Fig.~\ref{cachingwithmobility}, non-cooperative caching performs better than random caching, because BSs of non-cooperative caching store the most populary contents.

Fig.~\ref{automatagididedqlearning} shows the iteration curves of LAQL and $\varepsilon $-greedy Q-learning. There are 10000 iterations in each trial. As can be seen from the figure, the curves of Q-learning fluctuate a lot, and with the increase of $\epsilon$, Q-learning takes more trials to converge. Also can be seen from Fig.~\ref{automatagididedqlearning}, that the convergence of LAQL outperforms $\varepsilon $-greedy Q-learning eventually.

\section{Conclusion}\label{section:Conclusion}

In this article, content cooperative caching scheme were studied along with user mobility and content popularity prediction for wireless communication system. In order to fulfill backhaul constraints, an optimization problem of users' MOS maximization was formulated. 1) For user position and content popularity prediction, an SFBC-NN algorithm was proposed that utilized the powerful nonlinear fitting ability of SFBC-NN. The efficiency of the proposed scheme was validated by practical testing data set. 2) For content cooperative caching problem, a LAQL based cooperative caching algorithm was proposed, utilizing LA for optimal action selection in every state. The effectiveness of the proposed solution was illustrated by numerical experiments.

\section*{Appendix~A: Proof of Lemma~\ref{theorem:optimallemma2}} \label{Appendix:A}

The express of the objective function in (\ref{optimizationproblem}) is (\ref{MOSall}). To prove that the optimization problem in (\ref{optimizationproblem}) is NP-hard. We first consider the NP-hardness of the following problem.

\begin{myPro1}[$N$-Disjoint Set Cover Problem]\label{Def.2DSCP}

Given a set $\mathcal{A}$ and a collection $\mathcal{B}$, the $N$-disjoint set cover problem is to determine whether $\mathcal{B}$ can be divided into $N$ disjoint set covers or not. More generally, consider a bipartite graph $\mathcal{G} = \left( {\mathcal{A},\mathcal{B},\mathcal{E}} \right)$ with edges $\mathcal{E}$ between two disjoint vertex sets $\mathcal{A}$ and $\mathcal{B}$. Thus the $N$-disjoint set cover problem is whether there exist $N$ disjoint sets ${{\cal B}_1},{{\cal B}_2}, \cdots {{\cal B}_N} \subset {\cal B}$ in which $\left| {{{\cal B}_1}} \right| + \left| {{{\cal B}_2}} \right| +  \cdots  + \left| {{{\cal B}_N}} \right| = \left| {\cal B} \right|$ and ${\cal A} = \mathop U\limits_{\textbf{B} \in {{\cal B}_1}} N\left( \textbf{B} \right) = \mathop U\limits_{\textbf{B} \in {{\cal B}_2}} N\left( \textbf{B} \right) =  \cdots  = \mathop U\limits_{\textbf{B} \in {{\cal B}_N}} N\left( \textbf{B} \right)$. The $N$-disjoint set cover problem can be denoted as $\textbf{\textup{NDSC}}\left( {{\cal A},{\cal B},{\cal E}} \right)$.

\end{myPro1}

According to \cite{Cardei2005}, the $\textbf{NDSC}\left( {{\cal A},{\cal B},{\cal E}} \right)$ is proved to be NP-complete when $\left( {N \ge 2} \right)$. Consider $\mathcal{A}$ to be the set of mobile users, $\mathcal{B}$ to be BSs with caching capacity, and $\mathcal{E}$ to be the edges representing connections between mobile users and BSs. The content popularity is $P = \left[ {{p_1},{p_2}, \cdots ,{p_F}} \right]$ where ${p_f} < 1$ for $\forall f \in \mathcal{F}$. Without loss of generality, we set the caching capacity of each BS is equals to 1, which means that each BS can cache only one content. According to (\ref{MOSall}), the maximization of the total MOS of users can only be gained by the BS caches all the contents that the user requested. In other words, each user has at least one neighboring BS caching content 1 and other neighboring BS caching contents 2 to $F$. Letting ${{{\cal B}_1}}$, ${{{\cal B}_2}}$, and ${{{\cal B}_F}}$ be the disjoint sets of BSs containing file 1 to file $F$, separately. We conclude that determining whether the objective function (\ref{optimizationproblem}) is maximized is equivalent to determining the existence of and forming a $N$-disjoint set cover problem. From the analysis above, one can conclude that problem (\ref{optimizationproblem}) is NP-hard.

\section*{Appendix~B: Proof of Lemma~\ref{lemma:exclude_lieral}} \label{Appendix:B}
\begin{figure} [t!]
\centering
\includegraphics[width=3in]{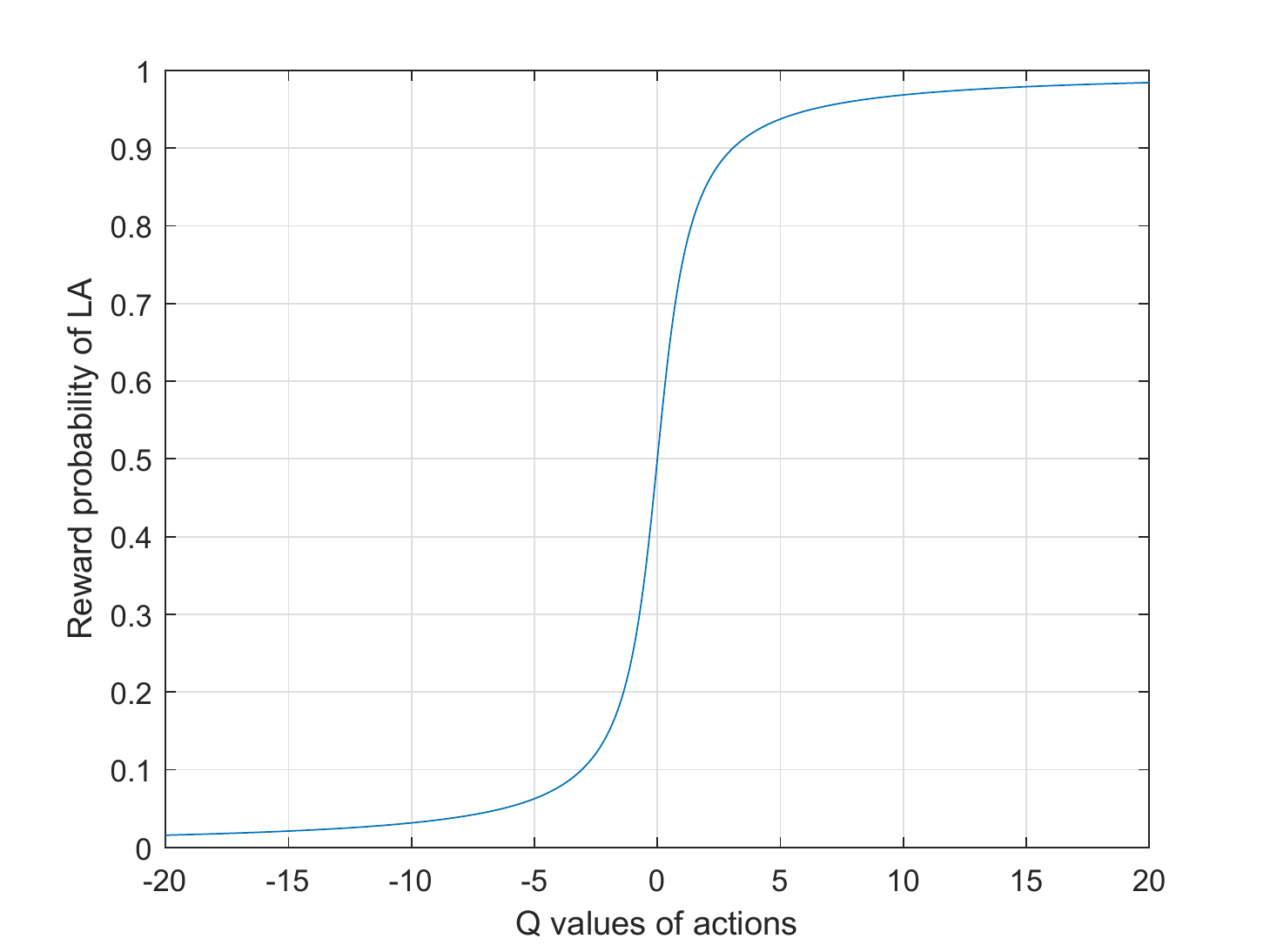}
 \caption{Mapping reward value to probability.}\label{tan}
\end{figure}

We shall prove \textbf{Lemma \ref{lemma:exclude_lieral}} in two steps. Firstly, we show that we can convert the average reward value for an action to a reward probability, given that a higher average reward corresponds a higher reward probability.  In other words, we need to map the value of Q value\footnote{For Q-learning, it is necessary to have bounded rewards for the actions in order to make the system converge. Here we use the range ($-\infty$, $+\infty$) to indicate the bounded rewards can be arbitrarily large.} from ($-\infty$, $+\infty$) to ($0, 1$). This conversion can be done by a modified arctangent-shape function, as shown in Fig. \ref{tan}.  Given that $Q(s,a)$ can converge to the optimal Q value, we can be certain that there exists a long term stationary reward behind each action of each state, and thus a stationary reward probability after conversion. Secondly, we need to show that the LA is $\epsilon$-optimal in the random and stationary environment, represented by the reward probability behind each action, but hidden before learning. The proof of $\epsilon$-optimal is not trivial and the detailed proof can be found in \cite{Zhang2016}. Here we just outline the main steps of the proof presently.

Now let's suppose there are $r$ actions for state $s$. The real reward in probability for action $i$, after conversion from the long term average reward value of action $i$, is $d_i$.  Correspondingly, the estimated reward probability by the LA for action $i$ at time $t$ is $\hat{d}(t)_i$, where $i\in\{1,\ldots, r\}$. Define $\bold{D}=[d_1, \ldots, d_r]$ and $\hat{\bold{D}}(t)=[\hat{d}_1(t), \ldots, \hat{d}_r(t)]$.  Let action probability vector be $\bold{P}(t)=[p_1(t), p_2(t), ..., p_r(t)]$, which is used to determine the issue of which action is to be selected at $t$, where $\sum\limits_{j=1...r}{p_j(t)}=1$.  Let $m$ be the index of the optimal action. To show the $\epsilon$-optimality of LA, we need to prove that given any $\epsilon > 0$ and $\delta >0$, there exist an $N_0>0$ and a $t_0 < \infty$ such that for all time $t \geq t_0$ and for any positive learning parameter $N>N_0$,
\begin{eqnarray}
Pr\{p_m(t) > 1 - \epsilon\} > 1-\delta.
\label{epslonoptimal}
\end{eqnarray}
This result has been proven in \cite{Zhang2016}, and we illustrate here only the outline of the proof, just for a quick reference to the theorem that has been used to support the result. One is recommended to refer to the original paper for a thorough elaboration of the reasoning.

The $\epsilon$-optimality of the DPA was based on the submartingale property of the Action Probability sequence $\{p_m(t)_{t>t_0}\}$. Thus the submartingale convergence theory and the theory of Regular functions were invoked to prove that the DPA will converge to the optimal action in probability, i.e.,
$Pr\{p_m(\infty)=1\} \rightarrow 1$.

$\{p_m(t)_{t>t_0}\}$ being a submartingale was proven by the definition of submartingale. Firstly, $p_m(t)$ is a probability, hence $E[p_m(t)] \leq 1< \infty$ holds. Secondly, according to the learning mechanism of DPA, the expectation of $p_m(t)$ can be calculated explicitly as:
\begin{align}
%& ~~~E[p_m(t+1)] \nonumber\\
&E[p_m(t+1)|\bold{P}(t)]  \nonumber\\
=& \sum\limits_{j=1...r}{p_j(t)\left(d_j\left(q(t)(p_m(t)+c_t\Delta)+(1-q(t))(p_m(t)-\Delta)\right)+(1-d_j)p_m(t)\right)} \nonumber \\
=& p_m(t) + \sum\limits_{j=1...r}{p_j(t)d_j\left(q(t)(c_t\Delta+\Delta)-\Delta\right)}, \nonumber
\end{align}
where $q(t)=Pr(\hat{d}_m(t)>\hat{d}_i(t))$, $\forall i \neq m$, is the accuracy probability of the reward estimates. By the definition of submartingale, one can see that if $ \forall t>t_0, ~~~ q(t) > \frac{\Delta}{c_t\Delta+\Delta}=\frac{1}{c_t+1} \geq \frac{1}{2}$ holds, then $E[p_m(t+1)|\bold{P}(t)]>p_m(t)$ becomes true, and $\{p_m(t)_{t>t_0}\}$ is thus a submartingale.

It has been proven in \cite{Zhang2016} that the accuracy probability of the reward estimates can be arbitrarily high. The proof itself involves quite a bit algebraic manipulations, and is thus complicated, but the rationale behind it is straight forward. It proved that after $t_0$, each action, with a very high probability, will be selected for a large number of times. Then, given that each action has been tried for a large number of times, the estimate of its reward will certainly become accurate enough to surpass the number $\frac{1}{2}$, which grantees $\{p_m(t)_{t>t_0}\}$ for being a submartingale.

Once the submartingale property is proven, by the submartingale convergence theory, $p_m(\infty) =0 \text{~or~} 1$ holds.
%More precisely, $P(\infty) =e_j, j = 1, 2, ..., \text{~or~} r$, where $e_j$ is the unit vector with the $j^{th}$ element being $1$.
To prove the DPA converges to the optional action, we need to shown that  $p_m(\infty) =1$ will happen with arbitrarily large probability, rather than $p_m(\infty) =0$. This is then equivalent to proving the convergence probability below
\begin{eqnarray}
\Gamma_m(\bold{P}) = Pr\{\bold{P}(\infty)=\bold{e}_m | \bold{P}(0) = \bold{P}\} \rightarrow 1,
%\text{~as~} t \rightarrow \infty. \nonumber
\label{converge}
\end{eqnarray}
where $\bold{e}_j$ is the unit vector with the $j^{th}$ element being $1$.
To prove Eq. (\ref{converge}), we now need to clarify the following definitions \cite{narendra2012learning}:

$\Phi(\bold{P})$: a function of $\bold{P}$.

$U$: an operator such that
$U\Phi(\bold{P})=E[\Phi(\bold{P}(n+1)) | \bold{P}(n)=\bold{P}],$\\
$~~~~~~~~~~~$applying $U$ for $n$ times:
$U^{n}\Phi(\bold{P})=E[\Phi(\bold{P}(n))|\bold{P}(0)=\bold{P}].$

The function $\Phi(\bold{P})$ is:
\begin{itemize}
\item
Superregular: If $U\Phi(\bold{P}) \leq \Phi(\bold{P})$.
Then applying $U$ repeatedly yields:
\begin{align}
\label{USuper}
\Phi(\bold{P})\geq U\Phi(\bold{P})\geq U^2\Phi(\bold{P}) \geq ... \geq U^{\infty}\Phi(\bold{P}).
\end{align}
\item
Subregular: If $U\Phi(\bold{P}) \geq \Phi(\bold{P})$.
In this case, if we apply $U$ repeatedly, we have
\begin{align}
\label{USub}
\Phi(\bold{P})\leq U\Phi(\bold{P})\leq U^2\Phi(\bold{P}) \leq ... \leq U^{\infty}\Phi(\bold{P}).
\end{align}
\item
Regular: If $U\Phi(\bold{P})=\Phi(\bold{P})$.
In such a case, it follows that:
\begin{align}
\label{URegu}
\Phi(\bold{P})= U\Phi(\bold{P})= U^2\Phi(\bold{P}) = ... = U^{\infty}\Phi(\bold{P}).
\end{align}
\end{itemize}

If we suppose $\Phi(\bold{P})$ satisfies the boundary conditions
$\Phi(\bold{e}_m)=1 \text{~and~} \Phi(\bold{e}_j)=0, (\forall j \neq m)$,
then, one can calculate $U^{\infty} \Phi(\bold{P})$ and obtain very interesting results as follows:
\begin{align}
\label{UGamma}
U^{\infty} \Phi(\bold{P}) &= E[\Phi(\bold{P}(\infty)) | \bold{P}(0)=\bold{P}]\nonumber \\
&=\sum_{j=1}^{r}{\Phi(\bold{e}_j)Pr\{\bold{P}(\infty)=\bold{e}_j | \bold{P}(0)=\bold{P} \}}, ~~\textit{~the ~definition~ of~ Expectation}\nonumber\\
&={\Phi(\bold{e}_m)Pr\{\bold{P}(\infty)=\bold{e}_m | \bold{P}(0)=\bold{P} \}}, ~~\textit{~the ~boundry~ condition $\Phi(\bold{e}_j)=0$,  $\forall~j \neq m$}\nonumber\\
&=Pr \{\bold{P}(\infty)=\bold{e}_m | \bold{P}(0)=\bold{P} \}, ~~\textit{~the ~boundry~ condition $\Phi(\bold{e}_m)=1$} \nonumber\\
&=\Gamma_m(\bold{P}).
%& U^{\infty}\phi_m(P)=\Gamma_m(P). \nonumber
\end{align}

This equation tells that the convergence probability can be studied by applying $U$ an infinite number of times to the function $\Phi(\bold{P})$. What's more, if $\Phi(\bold{P})$ is a Regular function, the sequence of operations will lead to a function that equals the function $\Phi(\bold{P})$ itself, whereas if $\Phi(\bold{P})$ is a subregular/superregular function, the sequence of operations can be bounded from below/above by $\Phi(\bold{P})$.

What the authors have done in \cite{Zhang2016}, is to find a subregular function of $\bold{P}$ that meets the boundary conditions, to bound the convergence probability of $\Gamma_m(\bold{P})$ from below. Then they proved the subregular function itself converges to unity, thus implies that the convergence probability, which is greater than or equal to the subregular function of $\bold{P}$, will converge to unity as well.

\vspace{-0.3cm}
\bibliographystyle{IEEEtran}
\bibliography{mybib}

\end{document}